\documentclass{article}

\usepackage{array}
\usepackage{listings}
\usepackage{fancyhdr}
\usepackage{lastpage}
\usepackage{float}
\usepackage{url}
\usepackage{xurl}
\usepackage{xcolor}
\usepackage{hyperref}
\usepackage{nameref}
\usepackage{adjustbox}
\usepackage{amssymb}
\usepackage{amsthm}
\usepackage{amsmath}
\usepackage{algorithm}
\usepackage{algpseudocode}
\usepackage{algcompatible}
\usepackage[numbers]{natbib}
\usepackage{tikz}
\usepackage{pgfplots}
\usepackage{textcomp}
\usepackage{subcaption}
\usepackage{fontawesome5}
\usepackage{tabularx}
\usepackage{multirow}
\usepackage[utf8]{inputenc}
\usepackage[T1]{fontenc}
\usepackage{graphicx}
\usepackage{ragged2e} 
\newcolumntype{Y}{>{\RaggedRight\arraybackslash}X}
\newcolumntype{L}{>{\RaggedRight\arraybackslash\bfseries}p{3cm}}
\usepackage[none]{hyphenat}
\hyphenchar\font=-1 
\usepackage[margin=1.2truein]{geometry} 
\hypersetup{
    colorlinks=true,
    linkcolor=red,
    urlcolor=red,
    citecolor=red
}

\usetikzlibrary{
    shapes,
    arrows,
    arrows.meta,
    positioning,
    backgrounds,
    shapes.multipart,
    calc
}

\tikzset{
    block/.style={draw, fill=white, rectangle, minimum height=48pt, minimum width=48pt},
    round/.style={draw, fill=white, circle},
    sum/.style={draw, fill=white, circle, node distance=2cm},
    input/.style={coordinate},
    output/.style={coordinate},
}

\pgfplotsset{compat=newest}

\definecolor{codegreen}{rgb}{0,0.6,0}
\definecolor{codegray}{rgb}{0.5,0.5,0.5}
\definecolor{codepurple}{rgb}{0.58,0,0.82}
\definecolor{backcolour}{rgb}{0.95,0.95,0.92}

\lstdefinelanguage{python}{
    basicstyle=\ttfamily\footnotesize,
    breaklines=true,                 
    captionpos=b,                    
    tabsize=2
}

\lstdefinelanguage{json}{
    basicstyle=\ttfamily\footnotesize,
    showstringspaces=false,
    keywordstyle=\color{blue},
    stringstyle=\color{black},
    string=[s]{"}{"}
}

\lstdefinestyle{sql}{
    language=SQL,
    basicstyle=\ttfamily\footnotesize,
    keywordstyle=\color{blue}\bfseries,
    identifierstyle=\color{black},
    stringstyle=\color{red},
    commentstyle=\color{green},
    morecomment=[l]{--},
    morecomment=[s]{/*}{*/}
}

\begin{document}
\title{
    Global Geolocated Realtime Data of\\Interfleet Urban Transit Bus Idling
}
\author{
    Nicholas Kunz \hyperlink{nhk37@cornell.edu}{nhk37@cornell.edu}\\
    H. Oliver Gao \hyperlink{hg55@cornell.edu}{hg55@cornell.edu}\\~\\
    Systems Engineering, Cornell University
}

\date{\today}
\maketitle
\thispagestyle{fancy}

\begin{abstract}
\noindent
Urban transit bus idling is a contributor to ecological stress, economic inefficiency, and medically hazardous health outcomes due to emissions. The global accumulation of this frequent pattern of undesirable driving behavior is enormous. To measure its scale, we propose GRD-TRT-BUF-4I (\textit{``Ground Truth Buffer for Idling''}), a realtime detection system that records the geolocation and idling duration of urban transit bus fleets internationally. Using live vehicle locations from General Transit Feed Specification (GTFS) Realtime, the system detects approximately 200,000 idling events per day from over 50 cities across North America, Europe, Oceania, and Asia. This realtime data supports operational decision-making and fleet management by enabling dynamic reduction of idling frequency and duration as events occur, while simultaneously capturing their cumulative effects. Civil and Transportation Engineers, Urban Planners, Epidemiologists, Policymakers, and others can use this data for emissions modeling, traffic management, route planning, and other urban sustainability efforts at a variety of geographic and temporal scales.\\

\noindent
\textbf{Keywords:} GTFS Realtime; vehicle idling detection; fleet telemetry; spatiotemporal event detection; open transit data; urban transit systems; urban sustainability\end{abstract}

\vspace{-12pt}
\begin{table}[H]
\centering
\begin{tabularx}{\textwidth}{|L|Y|}
\hline
Subject & Computer Sciences \\
\hline
Specific subject area &
Realtime idling event detection for worldwide urban transit bus fleet operations. \\
\hline
Type of data &
Realtime (\verb|.json|), Geospatial (\verb|.geojson|), Table (\verb|.csv|), Database (\verb|SQL|). \\
\hline
Data collection &
Live vehicle locations were collected from publicly accessible General Transit Feed Specification (GTFS) Realtime REST API endpoints. Data was ingested as Protocol Buffers at fixed intervals (default $r = 30$ seconds). An idling event detection system (GRD-TRT-BUF-4I) identified vehicles that remained geographically stationary across consecutive time steps beyond the parameterized threshold. All observations were recorded using WGS84 geospatial coordinates and Unix epoch timestamps.\\
\hline
Data source location &
Global: urban transit bus fleets across North America, Europe, Oceania, and Asia, including New York, Philadelphia, Washington D.C., Boston, Pittsburgh; Los Angeles, San Francisco, San Diego, Seattle, Sacramento, Portland; Atlanta, Miami, Tampa, Louisville, Nashville; Minneapolis--St. Paul, St. Louis, Madison, Columbus, Des Moines; Denver, Phoenix, San Antonio, Austin, Billings; Montréal, York, Hamilton, Halifax, Thunder Bay; Vancouver, Calgary, Edmonton, Saskatoon; Amsterdam, Stockholm, Helsinki, Dublin, Rome; Warsaw, Kraków, Gdańsk, Prague; Sydney, Brisbane, Adelaide, Auckland, Christchurch; and Delhi.\\
\hline
Data accessibility &
Repository name: Figshare \newline
Data identification number: \href{https://doi.org/10.6084/m9.figshare.25224224}{10.6084/m9.figshare.25224224} \newline
Direct URL to data: \href{https://doi.org/10.6084/m9.figshare.25224224}{https://doi.org/10.6084/m9.figshare.25224224} \newline
Instructions for accessing these data: Static (\verb|.csv|) files containing 24-hour global idling event records can be downloaded directly without authentication. Realtime access requires deployment of the open-source software and connection to GTFS Realtime endpoints. \\
\hline
Related research article &
None.\\
\hline
\end{tabularx}
\caption{Specifications Table}
\label{tab:specifications}
\end{table}

\vspace{-12pt}
\section{Value of the Data}
\label{sec:value}

\begin{itemize}
    \item This data establishes the first global, realtime, and geolocated record of urban transit bus idling events. It enables standardized comparison of idling behavior across cities, regions, and continents.
    
    \item The data supports high-resolution modeling of vehicle emissions, driving behavior, fuel consumption, and population exposure by preserving spatial and temporal structure at the level of individual idling events.

    \item Transit agencies and operators can use the data to identify persistent idling patterns and evaluate operational inefficiencies. The data supports immediate feedback and targeted interventions rather than only retrospective analyses.
    
    \item Civil and Transportation Engineers, Urban Planners, Epidemiologists, Policymakers, and others can use the data to compare policies and practices across transit agencies, municipalities, and broader jurisdictions.
    
    \item The data serves as a real-world benchmark for developing and validating urban spatiotemporal analytics, streaming detection algorithms, and geospatial data engineering under operational conditions on a global scale.
\end{itemize}

\section{Background}
\label{sec:back}

Urban transit buses provide essential transportation services in cities around the world. In Brazil, they are the most common mode of transportation \cite{ics2020}. In India, they are the third most preferred mode of transportation, where the size of their national fleet increased by nearly 30\% between 2009 and 2019, from nearly 118,00 vehicles to more than 152,000 \cite{ceew2019, india2019}. China also increased the size of its national fleet over roughly the same period, totaling nearly 600,000 vehicles in 2021 \cite{china2021}. In the United States, nearly half of all passenger trips on public transportation were facilitated by urban transit buses, amounting to approximately 6.2 billion trips in 2022, most of which were powered by diesel fuel \cite{apta2023a, apta2023b}.\vspace{12pt}

Data from the American Public Transportation Association (APTA) reported that in 2020, more than 96\% of all commuter buses in the United States were diesel-powered \cite{apta2023b}. All other fuel types such as compressed natural gas (CNG), liquefied natural gas (LNG), biodiesel, gasoline, propane, hydrogen, hybrid, and battery-electric accounted for less than 4\% of all fielded vehicles when combined \cite{apta2023b}. Even throughout Europe, where aggressive measures have been taken to transition away from diesel-powered vehicles, the most common type of newly fielded urban transit bus in 2021 was indeed diesel-powered \cite{icct2022}. Although many implications emerge in this regard, engine idling is a concern worldwide.\vspace{12pt}

Idling is when a vehicle’s engine is in operation, but is geographically stationary \cite{epa2006}. While individual idling events may seem insignificant, the total accumulation of these ephemeral, but repeated events is material. For instance, in 2022 the National Renewable Energy Laboratory (NREL) found that urban transit buses in the United States idled for roughly 40\% of their typical 9-hour operational period based on data from 16 vehicles over 19,440 hours (2.22 years) of cumulative driving time \cite{zhang2022}. An earlier study from 2010 reported similar findings throughout the United Kingdom, where idling rates were between 30\% and 44\% \cite{brightman2010}. A smaller scale, but more detailed example in South Korea found similar idling rates near 43\% \cite{park2019}.\vspace{12pt}

Ecological stress, economic inefficiency, and medically hazardous health outcomes - among others - all simultaneously emerge from this frequent and repeated pattern of undesirable driving behavior \cite{ziring2010, smit2020, fta2018}. A single idling urban transit bus can consume 0.5 gal (1.9 L) of diesel fuel per hour, emitting approximately 11.1 lbs (5 kg) of $\text{CO}_{2}$ equivalent GHG, 2.7 g of VOC, 2.7 g of THC, 37.4 g of CO, 61.1 g of $\text{NO}_{x}$, 1.2 g of $\text{PM}_{10}$, and 1.1 g of $\text{PM}_{2.5}$ \cite{kahn2006, epa2008}. These pollutants pose serious health risks including asthma, cardiovascular disease, liver and kidney damage \cite{kagawa2002, ito2011, suleimani2017}. Furthermore, noise from urban transit bus idling is a reported burden loud enough to potentially cause permanent hearing loss at SEL and $L_{max}$ A-weighted decibel levels of 111 dBA and 75 dBA, respectively \cite{smit2020, fta2018}.\vspace{12pt}

While the negative impacts of urban transit bus idling are numerous and evident, the magnitude of the problem has begun to emerge as a category of its own concern \cite{schipper2002, kujala2018}. Existing studies tend to focus on historical information from selected localities that are limited to a municipal, regional, or national fleet \cite{ziring2010, shan2019, sharma2019}. Few if any comprehensively measure the \textit{interfleet} phenomena in realtime on an international scale. Well-established developments like the General Transit Feed Specification (GTFS) Realtime have enabled transit agencies to share realtime data on service disruptions, vehicle locations, and arrival times using a standard schema since 2011 \cite{google2023, gtfs2020}. Growing international adoption of GTFS Realtime now affords the possibility to integrate these data sources using a common ingestion method \cite{kujala2018} as exhibited here.\vspace{12pt}

This geolocated realtime data is the first record of urban transit bus idling measured on a global scale. It immediately describes the worldwide phenomena \textit{when} and \textit{where} it occurs, giving transit agencies around the world the collective ability to actively intercept idling events, rather than merely measure the problem, only to analyze its negative impact after the fact. This shift would mark a significant advancement in the approach to reducing urban transit bus fleet idling worldwide by moving from reactive policies based on historical trends to preemptive measures based on proactive reduction strategies. The extensibility of this effort also allows easy integration of new GTFS Realtime sources as they become available.\vspace{12pt}

\section{Data Description}
\label{sec:data}

There are three ways to use the data. The first is through the \nameref{sub-sec:rlt-rsp}. The second is through a historical record of realtime responses in the \nameref{sub-sec:rel-dat}. The third is through the \nameref{sub-sec:stc-fil} queried from the \nameref{sub-sec:rel-dat}.

\subsection{Realtime Responses}
\label{sub-sec:rlt-rsp}

The realtime responses are JavaScript Object Notation (JSON) objects generated by the Subset microservice. Connecting to any of the Subset API's requires access to the region's REST API endpoint via Websocket. This is explained later in \nameref{sub-sec:sys-arc-arc} and \nameref{sec:notes}.\\

The JSON response will typically contain an array of key-value pairs. In the rare case that no idling events are detected, an empty response is returned without error. A typical JSON response contains eight keys: \textbf{iata\_id, vehicle\_id, route\_id, trip\_id, latitude, longitude, datetime,} and \textbf{duration}.\\

The JSON keys are equivalent to the fields found in \nameref{sub-sec:evt-tab}. Each field is described in Table \ref{tab:met-dat-evt} in the following section \nameref{sub-sec:rel-dat}. An example of a typical JSON response is exhibited in Figure \ref{fig:jsn-smp}.

\begin{figure}[H]
    \begin{adjustbox}{center}
        \begin{lstlisting}[language=json]
   [
      {
        "iata_id": "NYC",
        "vehicle_id": "MTA NYCT_9750",
        "route_id": "M42",
        "trip_id": "MQ_D3-Weekday-SDon-012900_M42_301",
        "latitude": 40.7625617980957,
        "longitude": -74.00098419189453,
        "datetime": 1697178720,
        "duration": 90
      },
      {
        "iata_id": "NYC",
        "vehicle_id": "MTA NYCT_9890",
        "route_id": "M104",
        "trip_id": "MV_D3-Weekday-SDon-011000_M104_101",
        "latitude": 40.814937591552734,
        "longitude": -73.95511627197266,
        "datetime": 1697178722,
        "duration": 120
      }
   ]
        \end{lstlisting}
    \end{adjustbox}
\caption{Example of Realtime Websocket Response as JSON Object}
\label{fig:jsn-smp}
\end{figure}

\break
\subsection{Relational Database}
\label{sub-sec:rel-dat}

The relational database is based on a common Dimensional Fact Model (DFM) implemented in PostgreSQL. It stores two tables. The first is the \textbf{Agency} table, which acts as the ``dimension table'', containing static and descriptive attributes. The second is the \textbf{Events} table, which acts as the ``fact table'' that stores the realtime idling events. The two are joined on the common field \textbf{iata\_id} or ``natural key'' as the primary key, exhibited in Figure \ref{fig:rel-dat}.\\

\begin{figure}[H]
\begin{center}
    \begin{tikzpicture}[thick, every node/.style={align=center}, >=Latex]
        \node[rectangle split, rectangle split parts=2, draw, text width=3.5cm] (agency) {
            \textbf{Agency}
            \nodepart[align=left]{two} \textbf{iata\_id}\\ agency\\ city\\ country\\ region \\ continent
        };
        \node[rectangle split, rectangle split parts=2, draw, text width=3.5cm, below right=of agency, yshift=1cm] (events) {
            \textbf{Events}
            \nodepart[align=left]{two} \textbf{iata\_id}\\ vehicle\_id\\ route\_id\\ trip\_id\\ latitude\\ longitude\\ datetime\\ duration
        };
        \draw[->] (agency.south) |- (events.west);
    \end{tikzpicture}
\caption{DFM Relational Schema Between \textbf{Agency} Table and \textbf{Events} Table}
\label{fig:rel-dat}
\end{center}
\end{figure}

\vspace{-18pt}
\subsubsection{Agency Table}
\label{sub-sec:agy-tab}

The \textbf{Agency} table contains six fields: \textbf{iata\_id, agency, city, country, region, continent}. Table \ref{tab:met-dat-agy} describes each field in greater detail. Table \ref{tab:smp-dat-agy} provides a truncated example taken from the region, US West. This is the ``dimension table'' in the DFM mentioned in \nameref{sub-sec:rel-dat}.

\begin{table}[H]
\begin{center}
    \resizebox{\textwidth}{!}{
        \begin{tabular}{|m{0.10\textwidth}|m{0.85\textwidth}|}
        \hline
        \textbf{Field} & \textbf{Description} \\
        \hline
        \textbf{iata\_id} & IATA identifier. A unique three-letter code designating the fleet's location.\\
        \hline
        \textbf{agency} & The transit agency responsible for administering and operating the fleet.\\
        \hline
        \textbf{city} & The name of the city associated with the IATA identifier and transit agency.\\
        \hline
        \textbf{country} & The country where the city is located.\\
        \hline
        \textbf{region} & The geographic region where the city is located.\\
        \hline
        \textbf{continent} & The continent on which the country is located.\\
        \hline
        \end{tabular}
    }
\caption{Field Descriptions of Agency Table}
\label{tab:met-dat-agy}
\end{center}
\end{table}

\vspace{-12pt}
\begin{table}[H]
\begin{center}
    \resizebox{\textwidth}{!}{
        \begin{tabular}{|m{0.1\textwidth}|m{0.1\textwidth}|m{0.15\textwidth}|m{0.15\textwidth}|m{0.2\textwidth}|m{0.15\textwidth}|}
        \hline
        \textbf{iata\_id} & \textbf{agency} & \textbf{city} & \textbf{country} &
        \textbf{region} &
        \textbf{continent}\\
        \hline
        LAX & (Metro) & Los Angeles & United States & United States West & North America\\
        \hline
        SFO & (MTC) & San Francisco & United States & United States West & North America\\
        \hline
        \end{tabular}
    }
\caption{Example of Agency Table from US West}
\label{tab:smp-dat-agy}
\end{center}
\end{table}

\subsubsection{Events Table}
\label{sub-sec:evt-tab}

The \textbf{Events} table contains eight fields: \textbf{iata\_id, vehicle\_id, route\_id, trip\_id, latitude, longitude, datetime, duration}. They are the same as the keys found in the JSON object, previously mentioned in \nameref{sub-sec:rlt-rsp}. Table \ref{tab:met-dat-evt} describes each field in greater detail. Table \ref{tab:smp-dat-evt} provides a truncated example taken from the geographic region, US West \textcolor{cyan}{\textbullet} as an example. 
This is the ``fact table'' in the DFM mentioned in \nameref{sub-sec:rel-dat}.

\begin{table}[H]
\begin{center}
    \resizebox{\textwidth}{!}{
    \begin{tabular}{|l|l|}
        \hline
        \textbf{Field} & \textbf{Description} \\
        \hline
        \textbf{iata\_id} & IATA identifier. A unique three-letter code designates the fleet's location.\\
        \hline
        \textbf{vehicle\_id} & Non-standard identifier of a single vehicle within a fleet.\\
        \hline
        \textbf{route\_id} & Non-standard identifier of each vehicle's specified transit route.\\
        \hline
        \textbf{trip\_id} & Non-standard identifier of each vehicle's on-network trajectory.\\
        \hline
        \textbf{latitude} & Geocoordinate of the north–south position of an idling event (WGS84).\\
        \hline
        \textbf{longitude} & Geocoordinate of the east-west position of an idling event (WGS84).\\
        \hline
        \textbf{datetime} & Unix epoch timestamp designating the start date and time of an idling event.\\
        \hline
        \textbf{duration} & Number of seconds elapsed since the start of an idling event.\\
        \hline
    \end{tabular}
    }
\caption{Field Descriptions of Events Table}
\label{tab:met-dat-evt}
\end{center}
\end{table}

\vspace{-12pt}
\begin{table}[H]
\begin{center}
    \resizebox{\textwidth}{!}{
    \begin{tabular}{|l|l|l|l|l|l|l|l|}
        \hline
        \textbf{iata\_id} & \textbf{vehicle\_id} & \textbf{route\_id} & \textbf{trip\_id} &  \textbf{latitude} & \textbf{longitude} & \textbf{datetime} & \textbf{duration} \\
        \hline
        LAX & 8538 & 70016... & 16-13... & 34.08... & -118.38... & 1704067086 & 80 \\
        \hline
        SFO & 8742 & SF:11... & SF:23... & 37.73... & -122.43... & 1704067076 & 76 \\
        \hline
        SEA & 7235 & 53452... & 10025... & 47.72... & -122.29... & 1704066915 & 79 \\
        \hline
    \end{tabular}
    }
\caption{Example of Events Table from US West}
\label{tab:smp-dat-evt}
\end{center}
\end{table}

\vspace{-16pt}
\subsection{Static Files}
\label{sub-sec:stc-fil}

There are three static files that can be downloaded in Comma Separated Value (\verb|.csv|) format. Each file is a result from a table join between the \textbf{Agency} table and the \textbf{Events} table. Static files contain all fields found in both tables over a 24-hour collection period. The first file \verb|test-data-a.csv| was collected from December 31, 2023 00:01:30 UTC to January 1, 2024 00:01:30 UTC. The second file \verb|test-data-b.csv| was collected from January 4, 2024 01:30:30 UTC to January 5, 2024 01:30:30 UTC. The third file \verb|test-data-c.csv| was collected from January 10, 2024 16:05:30 UTC to January 11, 2024 16:05:30 UTC. They can be downloaded at: \href{https://doi.org/10.6084/m9.figshare.25224224}{https://doi.org/10.6084/m9.figshare.25224224}.\\

Figure \ref{fig:sql-qry} exhibits the query used to create the first static file from the relational database mentioned in \nameref{sub-sec:rel-dat}. The subsequent files were created using this query corresponding to their respective datetime ranges. Table \ref{tab:smp-csv-fle} provides a truncated example of the files. Again, they can be downloaded in \verb|.csv| format. They are the same \verb|.csv| files used for evaluating data quality in \nameref{sub-sec:tech} within \nameref{sec:meth}.

\begin{figure}[ht]
    \begin{adjustbox}{center}
        \begin{lstlisting}[style=sql]
SELECT agency.*, events.vehicle_id, 
       events.trip_id, events.route_id,
       events.latitude, events.longitude,
       events.datetime, events.duration 
FROM agency 
LEFT JOIN events ON agency.iata_id = events.iata_id
WHERE events.datetime
BETWEEN EXTRACT(EPOCH FROM TIMESTAMP '2023-12-31 00:01:30') 
AND EXTRACT(EPOCH FROM TIMESTAMP '2024-01-01 00:01:30');
        \end{lstlisting}
    \end{adjustbox}
\caption{PostgreSQL Query for Retrieving Static File(s)}
\label{fig:sql-qry}
\end{figure}

\begin{table}[H]
\begin{center}
    \resizebox{\textwidth}{!}{
    \begin{tabular}{|l|l|l|l|l|l|l|l|l|l|l|l|}
        \hline
        \textbf{iat..} & \textbf{age..} & \textbf{cit..} & \textbf{cou..} & \textbf{veh..} & \textbf{rou..} & \textbf{tri..} &  \textbf{lat..} & \textbf{lon..} & \textbf{dat..} & \textbf{dur..} \\
        \hline
        MIA & Mia.. & Mia.. & Uni.. & 21174 & 5639655 & 28437 & 25.6... & -80.4... & 1704066908 & 240 \\
        SYD & Tra.. & Syd.. & Aus.. & 43054... & 2036432 & 2459... & -33.8... & 151.1... & 1704066908 & 127 \\
        BNE & Tra.. & Bri.. & Aus.. & 73DD... & 261484... & 777-... & -28.1... & 153.5... & 1704066908 & 90 \\
        \hline
    \end{tabular}
    }
\caption{Example of Static File}
\label{tab:smp-csv-fle}
\end{center}
\end{table}

\section{Experimental Design, Materials and Methods}
\label{sec:meth}

\subsection{Data Sources}
\label{sub-sec:dat-src}
Live vehicle locations from on-network urban transit bus fleets from over 50 cities were collected using GTFS Realtime. Although there are hundreds of GTFS Realtime sources, those that were selected, satisfied a combination of immediate availability, geographic variety, sufficient fleet size, daily operational length, and active support of each endpoint. Note that transit agencies may not elect to adopt GTFS Realtime for reasons regarding development cost, technical expertise and maintenance, usage of existing systems, lack of political support or priority \cite{voulgaris2023, frick2020}.\\

To ensure authenticity and reproducibility, data from GTFS Realtime was collected from publicly accessible Representational State Transfer (REST) Application Programming Interface (API) endpoints provided by the transit agencies directly, rather than relying on private third-party aggregation. All data sources were consumed as Protocol Buffers, sometimes referred to as \textit{``protobufs”} \cite{pbd2023, pbr2023}. An example of the deserialized protobuf schema is exhibited in Figure \ref{fig:pb-smp-hdr} and Figure \ref{fig:pb-smp-msg}.\\

Each data source was categorized according to its geographic region, where each region contained between 1 to 6 sources depending on the availability and stability of the REST API. Table \ref{tab:src-us-east} through Table \ref{tab:src-asia} exhibit each data source and its corresponding region. Some regions were nested within national categories belonging to the same continent. In other cases, few enough data sources were available that the continent directly represented the region, such as Oceania and Asia.

\begin{figure}[H]
\centering
    \begin{tikzpicture}
    \node[anchor=south west, inner sep=0] (image) at (0,0) {\includegraphics[width=\linewidth]{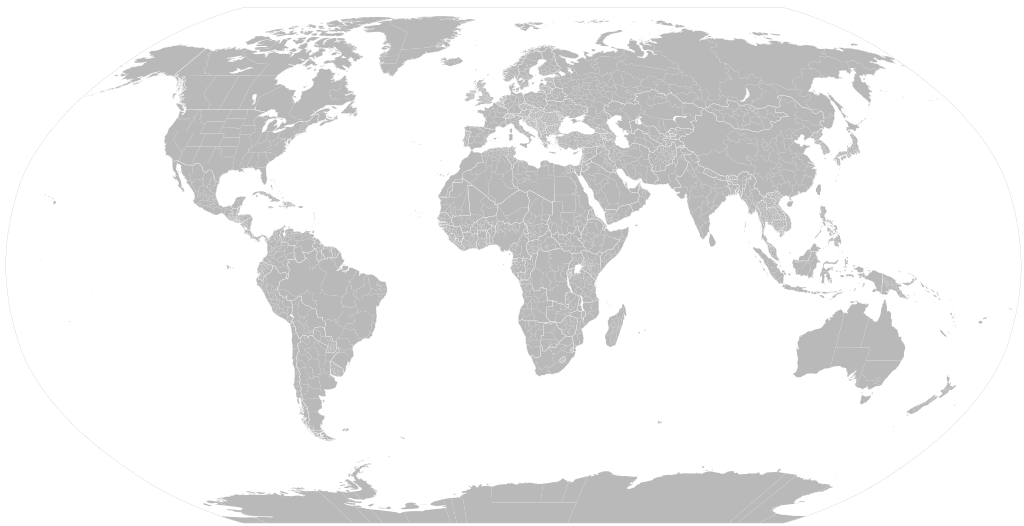}};
    \begin{scope}[x={(image.south east)}, y={(image.north west)}]

        \node at (0.25, 0.87) {North America};
        \node at (0.5, 0.91) {Europe};
        \node at (0.88, 0.365) {Oceania};
        \node at (0.74, 0.7) {Asia};
        
        \fill[magenta] (0.29,0.75) circle (2pt); 
        \fill[magenta] (0.285,0.745) circle (2pt); 
        \fill[magenta] (0.28,0.74) circle (2pt); 
        \fill[magenta] (0.3,0.755) circle (2pt); 
        \fill[magenta] (0.28,0.745) circle (2pt); 
        
        \fill[cyan] (0.16,0.725) circle (2pt); 
        \fill[cyan] (0.165,0.7) circle (2pt); 
        \fill[cyan] (0.178,0.785) circle (2pt); 
        \fill[cyan] (0.166,0.735) circle (2pt); 
        \fill[cyan] (0.173,0.77) circle (2pt); 

        \fill[brown] (0.255,0.695) circle (2pt); 
        \fill[brown] (0.26,0.655) circle (2pt); 
        \fill[brown] (0.258,0.66) circle (2pt); 
        \fill[brown] (0.255,0.73) circle (2pt); 
        \fill[brown] (0.25,0.72) circle (2pt);  

        \fill[orange] (0.249,0.77) circle (2pt); 
        \fill[orange] (0.242,0.754) circle (2pt); 
        \fill[orange] (0.242,0.74) circle (2pt);  
        \fill[orange] (0.251,0.765) circle (2pt); 
        \fill[orange] (0.266,0.74) circle (2pt);   

        \fill[yellow] (0.183,0.7) circle (2pt); 
        \fill[yellow] (0.209,0.735) circle (2pt); 
        \fill[yellow] (0.211,0.68) circle (2pt); 
        \fill[yellow] (0.215,0.685) circle (2pt);  

        \fill[teal] (0.295,0.77) circle (2pt); 
        \fill[teal] (0.28,0.765) circle (2pt); 
        \fill[teal] (0.279,0.764) circle (2pt); 
        \fill[teal] (0.32,0.771) circle (2pt); 
        \fill[teal] (0.26,0.79) circle (2pt); 
        
        \fill[black] (0.181,0.8) circle (2pt); 
        \fill[black] (0.209,0.81) circle (2pt);  
        \fill[black] (0.215,0.822) circle (2pt); 
        \fill[black] (0.229,0.81) circle (2pt);  
        \fill[black] (0.18,0.795) circle (2pt); 

        \fill[violet] (0.49,0.81) circle (2pt); 
        \fill[violet] (0.52,0.86) circle (2pt); 
        \fill[violet] (0.532,0.86) circle (2pt); 
        \fill[violet] (0.465,0.815) circle (2pt); 
        \fill[violet] (0.51,0.75) circle (2pt); 

        \fill[red] (0.527,0.812) circle (2pt); 
        \fill[red] (0.52,0.805) circle (2pt); 
        \fill[red] (0.512,0.8) circle (2pt);  
        \fill[red] (0.52,0.825) circle (2pt);  
        
        \fill[blue] (0.87,0.29) circle (2pt); 
        \fill[blue] (0.88,0.315) circle (2pt); 
        \fill[blue] (0.835,0.28) circle (2pt); 
        \fill[blue] (0.925,0.27) circle (2pt); 
        \fill[blue] (0.905,0.235) circle (2pt); 

        \fill[purple] (0.68,0.66) circle (2pt); 
        
    \end{scope}
    \end{tikzpicture}
\end{figure}

\begin{table}[H]
\centering
    \begin{tabularx}{\textwidth}{|X|X|X|}
    \hline
    \textbf{Continent} & \textbf{Country} & \textbf{Region} \\
    \hline
    \multirow{7}{*}{North America} & 
    \multirow{5}{*}{United States} & 
    US East \textcolor{magenta}{\textbullet} \\
    \cline{3-3}
    & & US West \textcolor{cyan}{\textbullet} \\
    \cline{3-3}
    & & US South \textcolor{brown}{\textbullet} \\
    \cline{3-3}
    & & US Central \textcolor{orange}{\textbullet} \\
    \cline{3-3}
    & & US Mountain \textcolor{yellow}{\textbullet} \\
    \cline{2-3}
    & \multirow{2}{*}{Canada} & 
    Canada East \textcolor{teal}{\textbullet} \\
    \cline{3-3}
    & & Canada West \textcolor{black}{\textbullet} \\
    \hline
    \multirow{2}{*}{Europe} & 
    \multirow{2}{*}{\textit{Greater}} & 
    EU West \textcolor{violet}{\textbullet}\\
    \cline{3-3}
    & & EU Central \textcolor{red}{\textbullet} \\
    \hline
    \multirow{1}{*}{Oceania} & 
    \multirow{1}{*}{\textit{Greater}} & 
    Oceania \textcolor{blue}{\textbullet} \\
    \hline
    \multirow{1}{*}{Asia} & 
    \multirow{1}{*}{\textit{Greater}} & 
    Asia \textcolor{purple}{\textbullet} \\
    \hline
    \end{tabularx}
\caption{GTFS Realtime Sources by Region}
\small
\textit{Greater} indicates the corresponding region contains more than one country.
\label{tab:src-reg}
\end{table}

Note that in Table \ref{tab:src-reg}, the \textit{Greater} designation in the Country column indicates the corresponding region contains more than one country. Similarly, the \textit{italicized} names in the City column in Table \ref{tab:src-us-east} through Table \ref{tab:src-asia} include neighboring cities from the broader metropolitan area. For example, \textit{Minneapolis} includes neighboring St. Paul, \textit{San Francisco} includes neighboring Oakland, etc. Other cities may also include locations beyond their proper municipal geographic boundary. However, those that are stylized were specified as such because of their immediate availability.
\vspace{12pt}

\begin{table}[H]
\centering
    \begin{tabularx}{\textwidth}{|>{\hsize=.11\hsize}X|>{\hsize=.14\hsize}X|>{\hsize=.66\hsize\raggedright\arraybackslash}X|>{\hsize=.04\hsize\arraybackslash}X|}
    \hline
    \textbf{Region} & \textbf{City} & \textbf{Agency} & \textbf{Ref.} \\
    \hline
    \multirow{5}{*}{US East \textcolor{magenta}{\textbullet}} & 
    New York & 
    Metropolitan Transportation Authority (MTA) & \cite{mta} \\
    \cline{2-4}
    & Philadelphia & 
    Southeastern Pennsylvania Transport. Authority (SEPTA) & \cite{septa}\\
    \cline{2-4}
    & \textit{Wash. D.C.} & 
    Washington Metropolitan Area Transit Authority (WMATA) & \cite{wmata} \\
    \cline{2-4}
    & Boston &
    Massachusetts Bay Transportation Authority (MBTA) & \cite{mbta} \\
    \cline{2-4}
    & Pittsburgh & 
    Pittsburgh Regional Transit (PRT) & \cite{prt} \\
    \hline
\end{tabularx}
\caption{US East GTFS Realtime Sources}
\small
\textit{Italicized} city names include the broader metropolitan area.
\label{tab:src-us-east}
\end{table}

\begin{table}[H]
\centering
    \begin{tabularx}{\textwidth}{|>{\hsize=.12\hsize}X|>{\hsize=.14\hsize}X|>{\hsize=.71\hsize\raggedright\arraybackslash}X|>{\hsize=.04\hsize\raggedright\arraybackslash}X|}
    \hline
    \textbf{Region} & \textbf{City} & \textbf{Agency} & \textbf{Ref.} \\
    \hline
    \multirow{6}{*}{US West \textcolor{cyan}{\textbullet}} & 
    Los Angeles & 
    Los Angeles County Metro. Transport. Authority (Metro) & \cite{la-metro} \\
    \cline{2-4}
    & \textit{San Fran.} & 
    Metropolitan Transportation Commission (MTC) & \cite{mtc} \\
    \cline{2-4}
    & San Diego &
    San Diego Metropolitan Transit System (MTS) & \cite{mts} \\
    \cline{2-4}
    & \textit{Seattle} &
    King County Metro (KCM) & \cite{kc-metro} \\
    \cline{2-4}
    & Sacramento & 
    Sacramento Regional Transit District (SacRT) & \cite{sacrt} \\
    \cline{2-4}
    & Portland & 
    Tri-County Metro. Transport. District of Oregon (TriMet) & \cite{trimet} \\
    \hline
    \end{tabularx}
\caption{US West GTFS Realtime Sources}
\small
\textit{Italicized} city names include the broader metropolitan area.
\end{table}

\begin{table}[H]
\centering
    \begin{tabularx}{\textwidth}{|>{\hsize=.12\hsize}X|>{\hsize=.12\hsize}X|>{\hsize=.66\hsize\raggedright\arraybackslash}X|>{\hsize=.04\hsize\raggedright\arraybackslash}X|}
    \hline
    \textbf{Region} & \textbf{City} & \textbf{Agency} & \textbf{Ref.} \\
    \hline
    \multirow{5}{*}{US South \textcolor{brown}{\textbullet}} & 
    Atlanta & 
    Metropolitan Atlanta Rapid Transit Authority (MARTA) & \cite{marta} \\
    \cline{2-4}
    & Miami & 
    Miami-Dade Transit (MDT) & \cite{mdt} \\
    \cline{2-4}
    & Tampa &
    Hillsborough Area Regional Transit (HART) & \cite{hart} \\
    \cline{2-4}
    & Louisville & 
    Transit Authority of River City (TARC) & \cite{tarc} \\
    \cline{2-4}
    & Nashville & 
    Nashville Metropolitan Transit Authority (Nashville MTA) & \cite{nas-mta} \\
    \hline
    \end{tabularx}
\caption{US South GTFS Realtime Sources}
\end{table}

\begin{table}[H]
\centering
    \begin{tabularx}{\textwidth}{|>{\hsize=.14\hsize}X|>{\hsize=.14\hsize}X|>{\hsize=.66\hsize\raggedright\arraybackslash}X|>{\hsize=.04\hsize\raggedright\arraybackslash}X|}
    \hline
    \textbf{Region} & \textbf{City} & \textbf{Agency} & \textbf{Ref.} \\
    \hline
    \multirow{5}{*}{US Central \textcolor{orange}{\textbullet}} & 
    \textit{Minneapolis} & 
    Metro Transit (Minnesota) & \cite{mn-metro} \\
    \cline{2-4}
    & St. Louis & 
    Metro Transit (St. Louis) & \cite{stl-metro} \\
    \cline{2-4}
    & Madison & 
    Metro Transit (Madison) & \cite{msn-metro} \\
    \cline{2-4}
    & Columbus &
    Central Ohio Transit Authority (COTA) & \cite{cota} \\
    \cline{2-4}
    & Des Moines & 
    Des Moines Area Regional Transit Authority (DART) & \cite{dart} \\
    \hline
    \end{tabularx}
\caption{US Central GTFS Realtime Sources}
\small
\textit{Italicized} city names include the broader metropolitan area.
\end{table}

\begin{table}[H]
\centering
    \begin{tabularx}{\textwidth}{|>{\hsize=.18\hsize}X|>{\hsize=.14\hsize}X|>{\hsize=.66\hsize\raggedright\arraybackslash}X|>{\hsize=.04\hsize\raggedright\arraybackslash}X|}
    \hline
    \textbf{Region} & \textbf{City} & \textbf{Agency} & \textbf{Ref.} \\
    \hline
    \multirow{5}{*}{US Mountain \textcolor{yellow}{\textbullet}} & 
    Denver & 
    Regional Transportation District (RTD) & \cite{rtd} \\
    \cline{2-4}
    & Phoenix & 
    Valley Metro Reg. Pub. Transport. Auth. (Valley Metro) & \cite{phx-metro} \\
    \cline{2-4}
    & San Antonio & 
    VIA Metropolitan Transit Authority (VIA Metro) & \cite{via-metro} \\
    \cline{2-4}
    & Austin & 
    Capital Metropolitan Transport. Authority (CapMetro) & \cite{cap} \\
    \cline{2-4}
    & Billings & 
    Billings Metropolitan Transit (MET) & \cite{met} \\
    \hline
    \end{tabularx}
\caption{US Mountain GTFS Realtime Sources}
\end{table}

\begin{table}[H]
\centering
    \begin{tabularx}{\textwidth}{|>{\hsize=.18\hsize}X|>{\hsize=.16\hsize}X|>{\hsize=.66\hsize\raggedright\arraybackslash}X|>{\hsize=.04\hsize\raggedright\arraybackslash}X|}
    \hline
    \textbf{Region} & \textbf{City} & \textbf{Agency} & \textbf{Ref.} \\
    \hline
    \multirow{5}{*}{Canada East \textcolor{teal}{\textbullet}} & 
    Montréal & Société de transport de Montréal (STM) & \cite{stm} \\
    \cline{2-4}
    & York & York Region Transit (YRT) & \cite{yrt} \\
    \cline{2-4}
    & Hamilton & Hamilton Street Railway (HSR) & \cite{hsr} \\
    \cline{2-4}
    & Halifax & Halifax Transit & \cite{hat} \\
    \cline{2-4}
    & Thunder Bay & Thunder Bay Transit & \cite{tbt} \\
    \hline
    \end{tabularx}
\caption{Canada East GTFS Realtime Sources}
\end{table}

\begin{table}[H]
\centering
    \begin{tabularx}{\textwidth}{|>{\hsize=.18\hsize}X|>{\hsize=.16\hsize}X|>{\hsize=.66\hsize\raggedright\arraybackslash}X|>{\hsize=.04\hsize\raggedright\arraybackslash}X|}
    \hline
    \textbf{Region} & \textbf{City} & \textbf{Agency} & \textbf{Ref.} \\
    \hline
    \multirow{4}{*}{Canada West \textcolor{black}{\textbullet}} & 
    Vancouver & TransLink (British Columbia) & \cite{tl-bc} \\
    \cline{2-4}
    & Calgary & Calgary Transit & \cite{cal} \\
    \cline{2-4}
    & Edmonton & Edmonton Transit System (ETS) & \cite{ets} \\
    \cline{2-4}
    & Saskatoon & Saskatoon Transit & \cite{sas} \\
    \hline
    \end{tabularx}
\caption{Canada West GTFS Realtime Sources}
\end{table}

\begin{table}[H]
\centering
    \begin{tabularx}{\textwidth}{|>{\hsize=.13\hsize}X|>{\hsize=.14\hsize}X|>{\hsize=.71\hsize\raggedright\arraybackslash}X|>{\hsize=.04\hsize\raggedright\arraybackslash}X|}
    \hline
    \textbf{Region} & \textbf{City} & \textbf{Agency} & \textbf{Ref.} \\
    \hline
    \multirow{5}{*}{EU West \textcolor{violet}{\textbullet}} & 
    \textit{Amsterdam} & Openbaar Vervoer (OV) & \cite{ov} \\
    \cline{2-4}
    & Stockholm & Storstockholms Lokaltrafik (SL) & \cite{sl} \\
    \cline{2-4}
    & Helsinki & Helsinki Regional Transport Authority (HSL) & \cite{hsl} \\
    \cline{2-4}
    & \textit{Dublin} & National Transport Authority & \cite{nta} \\
    \cline{2-4}
    & Rome & Azienda Tramvie e Autobus del Comune di Roma (ATAC) & \cite{atac} \\
    \hline
    \end{tabularx}
\caption{EU West GTFS Realtime Sources}
\small
\textit{Italicized} city names include the broader country.
\end{table}

\begin{table}[H]
\centering
\begin{tabularx}{\textwidth}{|>{\hsize=.16\hsize}X|>{\hsize=.12\hsize}X|>{\hsize=.71\hsize\raggedright\arraybackslash}X|>{\hsize=.04\hsize\raggedright\arraybackslash}X|}
    \hline
    \textbf{Region} & \textbf{City} & \textbf{Agency} & \textbf{Ref.} \\
    \hline
    \multirow{4}{*}{EU Central \textcolor{red}{\textbullet}} & 
    Warsaw & Zarząd Transportu Miejskiego w Warszawie (ZTM) & \cite{msw} \\
    \cline{2-4}
    & Krakow & Zarząd Transportu Publicznego w Krakowie (ZTP) & \cite{ztp} \\
    \cline{2-4}
    & Gdansk & Zarządu Transportu Miejskiego w Gdańsku (ZTM) & \cite{ztm} \\
    \cline{2-4}
    & Prague & Pražská Integrovaná Doprava (PID) & \cite{pid} \\
    \hline
\end{tabularx}
\caption{EU Central GTFS Realtime Sources}
\end{table}

\begin{table}[H]
\centering
    \begin{tabularx}{\textwidth}{|>{\hsize=.12\hsize}X|>{\hsize=.16\hsize}X|>{\hsize=.72\hsize\raggedright\arraybackslash}X|>{\hsize=.04\hsize\raggedright\arraybackslash}X|}
    \hline
    \textbf{Region} & \textbf{City} & \textbf{Agency} & \textbf{Ref.} \\
    \hline
    \multirow{5}{*}{Oceania \textcolor{blue}{\textbullet}} & 
    Sydney & Transport for New South Wales (Transport for NSW) & \cite{nsw} \\
    \cline{2-4}
    & Brisbane & Translink (Queensland) & \cite{tl-ql} \\
    \cline{2-4}
    & Adelaide & Adelaide Metro & \cite{adl-metro} \\
    \cline{2-4}
    & Auckland & Auckland Transport (AT) & \cite{at} \\
    \cline{2-4}
    & Christchurch & Environment Canterbury (Metro) & \cite{ecan} \\
    \hline
\end{tabularx}
\caption{Oceania GTFS Realtime Sources}
\end{table}

\begin{table}[H]
\centering
\begin{tabularx}{\textwidth}{|>{\hsize=.09\hsize}X|>{\hsize=.14\hsize}X|>{\hsize=.73\hsize\raggedright\arraybackslash}X|>{\hsize=.04\hsize\raggedright\arraybackslash}X|}
    \hline
    \textbf{Region} & \textbf{City} & \textbf{Agency} & \textbf{Ref.} \\
    \hline
    Asia \textcolor{purple}{\textbullet} & Delhi & Delhi Transport Corporation (DTC) & \cite{dtc} \\
    \hline
\end{tabularx}
\caption{Asia GTFS Realtime Sources}
\label{tab:src-asia}
\end{table}
\clearpage

\begin{figure}[H]
    \begin{adjustbox}{center}
        \begin{lstlisting}[language=json]
gtfs_realtime_version: "2.0"
incrementality: FULL_DATASET
timestamp: 1706573497
        \end{lstlisting}
    \end{adjustbox}
\caption{Example of GTFS Realtime Protocol Buffer Header Object}
\label{fig:pb-smp-hdr}
\end{figure}

\begin{figure}[H]
    \begin{adjustbox}{center}
        \begin{lstlisting}[language=json]
[
    id: "y0811"
    vehicle {
        trip {
            trip_id: "60487628"
            route_id: "216"
            direction_id: 1
            start_time: "19:16:00"
            start_date: "20240129"
            schedule_relationship: SCHEDULED
        }
        vehicle {
            id: "y0811"
            label: "0811"
        }
        position {
            latitude: 42.2721062
            longitude: -70.9509277
            bearing: 0
        }
        current_stop_sequence: 1
        stop_id: "3265"
        current_status: STOPPED_AT
        timestamp: 1706573492
        occupancy_status: MANY_SEATS_AVAILABLE
        occupancy_percentage: 0
    }
]
        \end{lstlisting}
    \end{adjustbox}
\caption{Example of GTFS Realtime Protocol Buffer Message Object}
\label{fig:pb-smp-msg}
\end{figure}

\subsection{Buffering Procedure}
\label{sub-sec:buf-pro}

Each geographic region (US East \textcolor{magenta}{\textbullet}, US West \textcolor{cyan}{\textbullet}, etc.) previously mentioned in \nameref{sub-sec:dat-src}, represents an asynchronous feed, $\text{Feed}_{i}$, exhibited in Figure \ref{fig:buf-pro}. It is constructed from its corresponding GTFS Realtime sources. As $\text{Feed}_{i}$ enters the buffer, it is stored in memory as $d_i$ according to indices $d_0, d_1, d_2, $...$, d_{h+1}$ at time $t, t+1, t+2, $...$, t+h+1$, respectively.\vspace{12pt}

After the buffer is populated to length $T$, the buffering procedure is fully initialized. As new $\text{Feed}_{i}$ enters the buffer at time $t$ once initialized, $\text{Feed}_{i+h+2}$ is omitted beyond buffer length $T$. This step is repeated theoretically to $\infty$ or practically until the procedure is stopped.\vspace{12pt}

After initializing the buffer, as well as when $\text{Feed}_{i}$ enters it, the index $d_0$ is treated as set $A$, index $d_h$ treated as set $B$, and index $d_{h+1}$ treated as set $C$. In other words, sets $A, B, C$ are equivalent to buffer indices $d_0, d_h, d_{h+1}$, respectively. The subsetting procedure is fully initialized with sets $A, B, C$.\vspace{12pt}

$n(A \cap B)$ is then computed, such that $A^n$ and $B^n$ are of length $n$. $\mathcal{A}$ and $\mathcal{B}$ are also of length $n$ and are further subsets of $A$ and $B$, only containing elements $x$ and $y$, respectively, subset from attributes $j$. Similarly, elements $z$ in $\mathcal{C}$ are subset from the same attributes $j$. $\mathcal{H}_i$ is then computed from $\mathcal{A} \cap \mathcal{B}$. Finally, $\mathcal{Y} = \mathcal{H} \cap \mathcal{C}$. The following section \nameref{sub-sec:sub-pro} provides more detail regarding the time-series set approach.

\begin{center}
\begin{figure}[H]
\begin{tikzpicture}[scale=1.5]

\draw[-] (0,0) -- (7,0);

\foreach \x in {0,1,...,3}{\draw (\x cm, 3pt) -- (\x cm, -3pt);}
\foreach \x in {5,6,7}{\draw (\x cm, 3pt) -- (\x cm, -3pt);}
\node[fill=white,rotate=90,inner sep=-1.25pt,outer sep=0,anchor=center] at (4,0){$\approx$};

\node[anchor=north] at (0,-0.1) {$t$};
\node[anchor=north] at (1,-0.1) {$t+1$};
\node[anchor=north] at (2,-0.1) {$t+2$};
\node[anchor=north] at (3,-0.1) {$t+3$};
\node[anchor=north] at (5,-0.1) {$t+h$};
\node[anchor=north] at (6,-0.1) {$t+h+1$};
\node[anchor=north] at (7,-0.1) {$T$};

\draw (0,1.25) rectangle (1,1.6) node[pos=.5] {$A$};
\draw (5,1.25) rectangle (6,1.6) node[pos=.5] {$B$};
\draw (6,1.25) rectangle (7,1.6) node[pos=.5] {$C$};

\draw (0,0.35) rectangle (1,0.75) node[pos=.5] {$d_0$};
\draw (1,0.35) rectangle (2,0.75) node[pos=.5] {$d_1$};
\draw (2,0.35) rectangle (3,0.75) node[pos=.5] {$d_2$};
\draw (3,0.35) rectangle (4,0.75) node[pos=.5] {$d_3$};
\draw (4,0.35) rectangle (5,0.75) node[pos=.5] {$\ldots$};
\draw (5,0.35) rectangle (6,0.75) node[pos=.5] {$d_h$};
\draw (6,0.35) rectangle (7,0.75) node[pos=.5] {$d_{h+1}$};

\draw[->] (-0.75,0.55) -- (-0.1,0.55) node [xshift=-24pt, yshift=1pt, label=left:$\text{Feed}_i$]{};
\draw[->] (-0.75,0.55) -- (-0.1,0.55);
\draw[->] (7.1,0.55) -- (7.75,0.55) node [xshift=-3pt, yshift=1pt, label=right:$\text{Feed}_{i+h+2}$]{};;
\draw[->] (7.1,0.55) -- (7.75,0.55);

\draw[->] (0.5,0.75) -- (0.5,1.2);
\draw[->] (5.5,0.75) -- (5.5,1.2);
\draw[->] (6.5,0.75) -- (6.5,1.2);

\draw[decorate,decoration={brace,amplitude=8pt}] (0.5,1.75) -- (5.5,1.75)
    node[anchor=south,midway,above=11pt]{
        $n(A \cap B) = (A^n, B^n)$
    };
\draw[decorate,decoration={brace,amplitude=8pt}] (0.5,2.5) -- (5.5,2.5)
    node[anchor=south,midway,above=11pt]{
        $\mathcal{A} = \bigl\{[x_{(i,\,j)}]: x \in A^n \bigl\}$, 
        $\mathcal{B} = \bigl\{[y_{(i,\,j)}]: y \in B^n \bigl\}$
    };
\draw[decorate,decoration={brace,amplitude=8pt}] (0.5,3.25) -- (5.5,3.25)
    node[anchor=south,midway,above=11pt]{
        $\mathcal{H}_i = \mathcal{A} \cap \mathcal{B}$,
        $\mathcal{C} = \bigl\{[z_{(i,\,j)}]: z \in C^c \bigl\}$
    };
\draw[decorate,decoration={brace,amplitude=8pt}] (5.5,3.25) -- (6.5,3.25)
    node[anchor=south,midway,above=11pt]{
        $\mathcal{Y} = \mathcal{H} \cap \mathcal{C}$
    };
\draw[decorate,decoration={brace,amplitude=6pt}] (7,-0.6) -- (0,-0.6)
    node[anchor=north,midway,below=11pt]{
        $T$ = Buffer Length
    };
\end{tikzpicture}
\caption{GTFS Realtime Buffering \& Subsetting Procedure}
\label{fig:buf-pro}
\end{figure}
\end{center}

\subsection{Subsetting Procedure}
\label{sub-sec:sub-pro}

Recall from \nameref{sub-sec:buf-pro} that $A$ is a set, and assume $a$ is a positive integer denoting the tuples whose elements $x_{(i,j)}$ at time $t$ belong to $A$, denoted as $A^a$, such that:

\begin{equation}
    A^a = \{x_{(i,j)\,t}: \text{for all} \, x \in A\}
\end{equation}\\
when expanded by $i$:
\begin{equation}
    A^a = \Biggl\{
        \begin{bmatrix}
            x_{(1,j)\,t}\\
            x_{(2,j)\,t}\\
            \vdots\\
            x_{(a,j)\,t}
        \end{bmatrix}: \text{for all} \, x \in A \Biggl\}
\end{equation}\\
and further by $j$:

\begin{equation}
    A^a = \bigl\{
        \begin{bmatrix}
            x_{(i, j=1,2,\ldots,m)\,t}\\
        \end{bmatrix}: \text{for all} \, x \in A \bigl\}
\end{equation}\\
where:
\begin{equation}
\begin{split}
    i & = \text{index}\\
    j & = \text{attribute}\\
\end{split}
\end{equation}\\
and $m$ is a positive integer denoting the attribute size.\\

Similarly, $B$ is a set, and $b$ is a positive integer denoting the tuples whose elements $y_{(i,j)}$ at time $t+h$ belong to $B$, denoted as $B^b$, such that:
\begin{equation}
    B^b = \{y_{(i,j)\,t+h}: \text{for all} \, y \in B\}
\end{equation}
where:
\begin{equation}
\begin{split}
    h & = \text{time-horizon}\\
\end{split}
\end{equation}
The intersection of $A^a$ and $B^b$ is then computed as:
\begin{equation}
    n(A^a \cap B^b) = A^a \cap B^b
\end{equation}\\
where subsets $A^n$ and $B^n$ now have have equal cardinality and $n$ denotes the equal index length, such that $|A^n| = |B^n|$, where:

\begin{equation}
\begin{aligned}[t]
    A^n & = \{x_{(i,j)\,t}: x \in A\}\\
\end{aligned}
\qquad 
\begin{aligned}[t]
    B^n & = \{y_{(i,j)\,t+h}: y \in B\}
\end{aligned}
\end{equation}\\
when expanded by the $i^{th}$ index:
\begin{equation}
\begin{aligned}[t]
    A^n = \Biggl\{
        \begin{bmatrix}
            x_{(1,j)\,t}\\
            x_{(2,j)\,t}\\
            \vdots\\
            x_{(n,j)\,t}
        \end{bmatrix}: x \in A \Biggl\}
\end{aligned}
\qquad 
\begin{aligned}[t]
    B^n = \Biggl\{
        \begin{bmatrix}
            y_{(1,j)\,t+h}\\
            y_{(2,j)\,t+h}\\
            \vdots\\
            y_{(n,j)\,t+h}
        \end{bmatrix}: y \in B \Biggl\}
\end{aligned}
\end{equation}\\
Further denoting subsets $A^n \subseteq A^b$ and $B^n \subseteq B^b$, we have:
\begin{equation}
\begin{aligned}[t]
    \mathcal{A} & = \bigl\{
        \begin{bmatrix}
            x_{(i,j=1,\hdots,5)\,t}\\
        \end{bmatrix}: x \in A^n \bigl\}
\end{aligned}
\qquad 
\begin{aligned}[t]
    \mathcal{B} & = \bigl\{
        \begin{bmatrix}
            y_{(i,j=1,\hdots,5)\,t+h}\\
        \end{bmatrix}: y \in B^n \bigl\}
\end{aligned}
\end{equation}\\

so that attributes $j=1,\hdots,5$ in subsets $A^n$ and $B^n$ are referred to directly as $\mathcal{A}$ and $\mathcal{B}$ from $t$ to $t+h$, respectively, and where $j=1,\hdots,5$ is:
\begin{equation}
\begin{split}
    1 & = \texttt{vehicle\_id}\\
    2 & = \texttt{route\_id}\\
    3 & = \texttt{trip\_id}\\
    4 & = \texttt{latitude}\\
    5 & = \texttt{longitude}
\end{split}
\end{equation}\\
Next, the subset $\mathcal{H}$ is computed as the intersection of $\mathcal{A}$ and $\mathcal{B}$, where $\mathcal{H} \subseteq \mathcal{A}$ and $\mathcal{H} \subseteq \mathcal{B}$, whose elements $w_{(i,j)}$ at time $t+h$ belong to  $\mathcal{H}$, such that:
\begin{equation}
    \mathcal{H} = \bigl\{
        \begin{bmatrix}
            w_{(i; j)\,t+h}\\
        \end{bmatrix}: w \in \mathcal{A} \cap \mathcal{B} \}
\end{equation}\\
when expanded by both $i^{th}$ index and $j^{th}$ attributes:
\begin{equation}
    \mathcal{H}^{p} = \Biggl\{
        \begin{bmatrix}
            w_{(1; \, j=1,\hdots,5)\,t+h}\\
            w_{(2; \, j=1,\hdots,5)\,t+h}\\
            \vdots\\
            w_{(p; \, j=1,\hdots,5)\,t+h}
        \end{bmatrix}: w \in \mathcal{A} \cap \mathcal{B} \Biggl\}
\end{equation}\\

It is important to highlight that the subset $\mathcal{H}$ is treated as a special case whose elements $w_{(i,j)}$ can remain in the subset throughout any sequence of time-steps and is appended as $\mathcal{H}_i$ to length $\mathcal{H}^p$. This differs from $\mathcal{A}$ and $\mathcal{B}$ whose elements can change at every time-step.\\

Similar to original sets $A$ and $B$, let $C$ be a set, and $c$ be a positive integer denoting the tuples whose elements $z_{(i,j)}$ at time $t+h+1$ belong to set $C$, denoted as $C^c$. This differs from the convention that $c$ is the complement, such that:
\begin{equation}
    C^c = \{z_{(i,j)\,t+h+1}: \text{for all} \, z \in C \}
\end{equation}\\
when expanded by the $i^{th}$ index:
\begin{equation}
    C^c = \Biggl\{
        \begin{bmatrix}
            z_{(1,j)\,t+h+1}\\
            z_{(2,j)\,t+h+1}\\
            \vdots\\
            z_{(c,j)\,t+h+1}
        \end{bmatrix}: \text{for all} \, z \in C \Biggl\}
\end{equation}\\

To match the attributes in set $C$ to those found in subsets $\mathcal{A}$ and $\mathcal{B}$, let the following be true:
\begin{equation}
\begin{aligned}[t]
    \mathcal{C} & = \bigl\{
        \begin{bmatrix}
            _{(i,j=1,\hdots,5)\,t+h+1}\\
        \end{bmatrix}: z \in C \bigl\}
\end{aligned}
\end{equation}

so that attributes $j=1,\hdots,5$ in set $C$ are referred to directly as subset $\mathcal{C}$ at time $t+h+1$. Next, the intersection of $\mathcal{C}$ and $\mathcal{H}$ is computed, where:
\begin{equation}
    \mathcal{Y} = \mathcal{C} \cap \mathcal{H}
\end{equation}\\
$\mathcal{Y}$ is now a further subset of $\mathcal{C}$ and $\mathcal{H}$, $\mathcal{Y} \subseteq \mathcal{C}$ and $\mathcal{Y} \subseteq \mathcal{H}$, whose elements $v_{(i,j)}$ at time $T$ belong to $\mathcal{Y}$, such that:\\
\begin{equation}
    \mathcal{Y} = \bigl\{
        \begin{bmatrix}
            v_{(i; j)\,T}\\
        \end{bmatrix}: v \in \mathcal{C} \cap \mathcal{H} \}
\end{equation}\\
when expanded by both $i^{th}$ index and $j^{th}$ attributes:
\begin{equation}
    \mathcal{Y}^{k} = \Biggl\{
        \begin{bmatrix}
            v_{(1,j=1,\hdots,5)\,T}\\
            v_{(2,j=1,\hdots,5)\,T}\\
            \vdots\\
            v_{(k,j=1,\hdots,5)\,T}
        \end{bmatrix}: v \in \mathcal{C} \cap \mathcal{H} \Biggl\}
\end{equation}\\
Note that the subset $\mathcal{Y}$ is also treated as a special case whose elements $v_{(i,j)}$ can remain in the subset throughout any sequence of time-steps. However, it is \textit{not} appended, as it is the final subset of length $\mathcal{Y}^k$. As the final output, $\mathcal{Y}$ resolves this series of expressions describing the subsetting procedure. The following section \nameref{sub-sec:com-pro} describes its parameters and how it's computed.\\

\subsection{Computational Procedure} 
\label{sub-sec:com-pro}

Combining \nameref{sub-sec:buf-pro} and \nameref{sub-sec:sub-pro}, \textbf{Algorithm 1} GRD-TRT-BUF-4I casts them into step-by-step instructions for computing realtime idling events. Details of the algorithm are provided in \nameref{sub-sec:algo-info}. There are three parameters:

\begin{enumerate}
    \item $r$: the rate at which GTFS Realtime server requests are made (seconds). It is a positive integer. $r$ has a default value of 30 seconds and should not exceed the rate limit of any GTFS Realtime server. It is recommended that $r$ not take values less than the frequency of vehicle location updates, which are typically updated every 30 seconds.
    \item $h$: the number of time-steps $t$ after an idling event is measured (interval). It is a positive integer. $h$ has a default value of 1. In the case that the default value of $r$ remains unchanged, any number of seconds beyond 60 seconds is recorded as an idling event. Alternatively, specifying $h$ as 2 would consider an idling event any number of seconds beyond 90 seconds, all else equal.
    \item $m$: a constant used to bound the length of appended subset $\mathcal{H}$ (iterations). It is a positive integer. $m$ has an arbitrary default value of 10. It is the maximum allowable iterations that elements $z \in \mathcal{H}$ are stored before being omitted when also not found in set $\mathcal{C}$. In other words, the elements $z$ are removed from the appended set $\mathcal{H}$ after $m$ number of iterations they are not within $\mathcal{H} \cap C$.
\end{enumerate}

After $r$, $h$, and $m$ are specified, an infinite loop is conducted outside the process in which \nameref{sub-sec:buf-pro} and \nameref{sub-sec:sub-pro} are both initialized and computed. Additional instructions in GRD-TRT-BUF-4I from steps 10 to 22 correspond to the subset $\mathcal{H}$. Again, details of the algorithm are provided in \nameref{sub-sec:algo-info}.\\

Finally, $\mathcal{Y}$ is computed as the primary output and is immediately returned and stored on disk. The following section \nameref{sub-sec:sys-arc} broadly outlines a collection of components and their relationships in this regard.

\vspace{-6pt}
\subsection{System Architecture \& Design}
\label{sub-sec:sys-arc}

\subsubsection{System Architecture}
\label{sub-sec:sys-arc-arc}
The system contains the basic functions of \textbf{Algorithm 1} GRD-TRT-BUF-4I. It is modeled through the Integration Definition for Process Modeling (IDEF0), exhibited in Figure \ref{fig:sys-arc}. Each component takes an input (left pointer), yields an output (right pointer), and requires a mechanism (bottom pointer). Those that process intermediate results also take control parameters (top pointer).\\

First, the Serve component inputs ``Fleet’’ - the telemetry data collected from the live location of vehicles in the network of urban transit bus fleets. It outputs ``$\text{Feed}_{i}$'', the GTFS Realtime feeds, and requires the open internet or ``Web’’ as a mechanism. Note that this process is modeled for architectural context, where no explicit controls are required.\\

The Extract component inputs the GTFS Realtime sources, $\text{Feed}_{i}$ and outputs the extracted data $d_{i}$, requiring an API key(s) or ``Key'’ as a mechanism to authorize REST API endpoint(s) access where required. The request rate is controlled by $r$, as previously described in \nameref{sub-sec:com-pro}. The purpose of this component is to asynchronously extract GTFS Realtime data.\\

The Buffer component inputs extracted data $d_i$ and outputs sets $A$, $B$, $C$ taken from buffer indicies $d_0, d_h, $...$, d_{h+1}$, respectively. It uses RAM or ``Memory’’ as a mechanism to store the contents and is controlled by $h$, as previously described in \nameref{sub-sec:com-pro}. The purpose of this component is to roll the extracted data into and out of memory.\\

The Subset component inputs sets $A$, $B$, $C$ and outputs the realtime idling events $\mathcal{Y}$. It uses a series of set operations or ``Operators’’ as a mechanism, and is controlled by $m$, as previously described in \nameref{sub-sec:com-pro}. The purpose of this component is to compute the primary logic of \textbf{Algorithm 1} GRD-TRT-BUF-4I.\\

Finally, the Store component inputs the realtime idling events $\mathcal{Y}$ and outputs them as historical information ``Data’’. It uses a database or ``Disk'’ as a mechanism. The purpose of this component is to capture the historical record of realtime idling events so that they can be queried in downstream analyses. No explicit controls are required.

\begin{figure}[H]
\centering
\begin{tikzpicture}[auto, thick, >=triangle 45, scale=0.90, transform shape]

    \draw node at (-4.5, 1) [input, name=position]{};
    \draw node at (-3, -0.5) [input, name=network]{};
    \draw node at (-3, 1) [block, name=server, label={[xshift=21pt, yshift=-48pt]}]{Serve};
    
    \draw node at (0, 1.5) [input, name=rate]{};
    \draw node at (0, -1.5) [input, name=key]{};
    \draw node at (0, 0) [block, name=request, label={[xshift=21pt, yshift=-48pt]}]{Extract};

    \draw node at (3, 0.5) [input, name=h]{};
    \draw node at (3, -2.5) [input, name=memory]{};
    \draw node at (3, -1) [block, name=buffer, label={[xshift=21pt, yshift=-48pt]}]{Buffer};

    \draw node at (6, -0.5) [input, name=sort]{};
    \draw node at (6, -3.5) [input, name=object]{};
    \draw node at (8, -2) [input, name=data]{};
    \draw node at (6, -2) [block, name=subset, label={[xshift=21pt, yshift=-48pt]}]{Subset};

    \draw node at (9, -4.5) [input, name=disk]{};
    \draw node at (9, -3) [block, name=store, label={[xshift=21pt, yshift=-48pt]}]{Store};

    \draw node at (10.5, -3) [input, name=data]{};

    \begin{scope}[transform canvas={xshift=0pt}]
        \draw[->] (position) -- node [xshift=-6pt, yshift=-3pt, label=left:Fleet]{}(server);
    \end{scope}

    \begin{scope}[transform canvas={xshift=0pt}]
        \draw[->] (network) -- node [xshift=3pt, yshift=-6pt, label=below:Web]{}(server);
    \end{scope}
    
    \begin{scope}[transform canvas={xshift=0pt}]
        \draw[->] (server) -- +(1.5, 0) |- node [xshift=2pt, yshift=-2pt, label=below: $\text{Feed}_i$]{}(request);
    \end{scope}
    
    \begin{scope}[transform canvas={xshift=0pt}]
        \draw[->] (rate) -- node [xshift=-4pt, yshift=6pt, label=above:Rate $(r)$]{}(request);
    \end{scope}

    \begin{scope}[transform canvas={xshift=0pt}]
        \draw[->] (key) -- node [xshift=3pt, yshift=-6pt, label=below:Key]{}(request);
    \end{scope}

    \begin{scope}[transform canvas={yshift=0pt}]
        \draw[->] (request) -- +(1.5, 0) |- node [xshift=0pt, yshift=-2pt, label=below:$d_i$]{}(buffer);
    \end{scope}

    \begin{scope}[transform canvas={xshift=0pt}]
        \draw[->] (h) -- node [xshift=-4pt, yshift=6pt, label=above:Time $(h)$]{}(buffer);
    \end{scope}    

    \begin{scope}[transform canvas={xshift=0pt}]
        \draw[->] (memory) -- node [xshift=3pt, yshift=-6pt, label=below:Memory]{}(buffer);
    \end{scope}

    \begin{scope}[transform canvas={yshift=0pt}]
        \draw[->] (buffer) -- +(1.5, 0) |- node [xshift=-4pt, yshift=-2pt, label=below:$ABC$]{}(subset);
    \end{scope}

    \begin{scope}[transform canvas={yshift=0pt}]
        \draw[->] (sort) -- node [yshift=6pt, label=above:Count $(m)$]{}(subset);
    \end{scope}
    
    \begin{scope}[transform canvas={xshift=0pt}]
        \draw[->] (object) -- node [xshift=3pt, yshift=-6pt, label=below:Operators]{}(subset);
    \end{scope}

    \begin{scope}[transform canvas={yshift=0pt}]
        \draw[->] (subset) -- +(1.5, 0) |- node [xshift=-4pt, yshift=-2pt, label=below:$\mathcal{Y}$]{}(store);
    \end{scope}

    \begin{scope}[transform canvas={xshift=0pt}]
        \draw[->] (disk) -- node [xshift=3pt, yshift=-6pt, label=below:Disk]{}(store);
    \end{scope}

    \begin{scope}[transform canvas={xshift=0pt}]
        \draw[->] (store) -- node [xshift=42pt, yshift=-4pt, label=left:Data]{}(data);
    \end{scope}

    \draw[dotted] (-2, -4.5) rectangle (7.75, 2.5);
    \node at (0.9, -4.1) {\textbf{Algorithm 1} GRD-TRT-BUF-4I};
    \end{tikzpicture}\\~\\
\caption{IDEF0 System Architecture of GRD-TRT-BUF-4I Algorithm}
\label{fig:sys-arc}
\end{figure}

\subsubsection{System Design}
\label{sub-sec:sys-des}

The system design implements a more detailed construction of the \nameref{sub-sec:sys-arc-arc}. It casts the IDEF0 architecture into a common \textit{Extract, Transform, Load} (ETL) microservice design pattern based on geographic region (US East \textcolor{magenta}{\textbullet}, US West \textcolor{cyan}{\textbullet}, etc.), exhibited in Figure \ref{fig:sys-des}. The system uses templated ETL pipelines per region, where each microservice corresponds to the ETL phase that it is nested within. Each microservice is then duplicated an arbitrary number of times, increasing redundancy to avoid single points of failure.\\

For each geographic region (\textbf{Region 1, Region 2, $...$, Region $N$}), the \textit{Extract} phase asynchronously extracts data from the associated GTFS Realtime servers (Server 1, Server 2, $...$, Server $n$), previously outlined in \nameref{sub-sec:sys-arc-arc}. The \textit{Transform} phase then computes GRD-TRT-BUF-4I from the extracted data, as previously mentioned in \nameref{sub-sec:com-pro}. Finally, the \textit{Load} phase uses the ``Write’’ microservices to insert the realtime idling event data into the Events table within the ``Store’’ microservice.\\

Modular decoupling of components into microservices allows users to directly access any phase of the ETL pipeline throughout any and all geographic regions. The design not only allows this, but strongly encourages it. To use the realtime idling event data, users access the ``Subset'' microservice directly, rather than through ``Write'' or ``Store''. This is achieved through the Subset API's and is explained in \nameref{sec:notes}.\\
\clearpage

Note that while casting the IDEF0 architecture into an ETL microservice design, the ``Buffer'' components are integrated within the ``Subset'' microservices along with parameters \textit{r}, \textit{h}, and \textit{m} as a practical design choice. The ``Write'' microservices are introduced to appropriately decouple their associated function per geographic region, whereas the alternative would require a tightly coupled design with either ``Subset'' or ``Store''. In addition, the Agency table is persisted within ``Store’’ and is explained in \nameref{sec:data}.\\

\begin{figure}[H]
\centering
\begin{tikzpicture}[auto, thick, >=triangle 45, scale=0.90, transform shape]

    \draw[fill=gray!20, draw=none, rounded corners] (-1.25, 2) rectangle (1, -7.25);
    \draw[fill=gray!20, draw=none, rounded corners] (1.75, 2) rectangle (4, -7.25);
    \draw[fill=gray!20, draw=none, rounded corners] (4.75, 2) rectangle (7.15, -7.25);
    \node at (-0.1, 1.6) {\textit{Extract}};
    \node at (2.9, 1.6) {\textit{Transform}};
    \node at (5.91, 1.6) {\textit{Load}};

    \node at (11.5, -3) [name=users] {\textbf{User}}; 

    \draw node at (-3, 0.6) [block, minimum height=10pt, name=server1]{Server 1};
    \draw node at (-3, 0) [block, minimum height=10pt, name=server2]{Server 2};
    \draw node at (-3, -0.6) [block, minimum height=10pt, name=server3]{Server $n$};

    \draw node at (-0.2, 0.2) [block]{};
    \draw node at (-0.1, 0.1) [block]{};
    \draw node at (-0.72, 0.6) [name=request1-1]{};          
    \draw node at (0, 0) [block, name=request1-2]{Extract};  
    \draw node at (-0.72, -0.6) [name=request1-3]{};         

    \draw node at (2.8, 0.2) [block]{};
    \draw node at (2.9, 0.1) [block]{};
    \draw node at (3, 0) [block, name=subset, label={[yshift=12pt]center:Subset}]{};
    \draw node at (3, -0.33) [block, minimum height=20pt, minimum width=40pt, name=buffer]{Buffer};

    \draw[dashed, ->] ([yshift=0.3cm]subset.east) -| ([yshift=0.3cm]users.north);  

    \draw node at (5.8, 0.2) [block]{};
    \draw node at (5.9, 0.1) [block]{};
    \draw node at (6, 0) [block, name=write]{Write};
    
    \draw node at (-3, -2.4) [block, minimum height=10pt, name=server4]{Server 1};
    \draw node at (-3, -3) [block, minimum height=10pt, name=server5]{Server 2};
    \draw node at (-3, -3.6) [block, minimum height=10pt, name=server6]{Server $n$};

    \draw node at (-0.2, -2.8) [block]{};
    \draw node at (-0.1, -2.9) [block]{};
    \draw node at (-0.72, -2.4) [name=request2-1]{};          
    \draw node at (0, -3) [block, name=request2-2]{Extract};
    \draw node at (-0.72, -3.6) [name=request2-3]{};          

    \draw node at (2.8, -2.8) [block]{};
    \draw node at (2.9, -2.9) [block]{};
    \draw node at (3, -3) [block, name=subset2, label={[yshift=12pt]center:Subset}]{};
    \draw node at (3, -3.33) [block, minimum height=20pt, minimum width=40pt, name=buffer2]{Buffer};

    \draw[dashed, ->] ([yshift=0.3cm]subset2.east) -- ([yshift=0.3cm]users.west);  
    
    \draw node at (5.8, -2.8) [block]{};
    \draw node at (5.9, -2.9) [block]{};
    \draw node at (6, -3) [block, name=write2]{Write};
    
    \draw node at (-3, -5.5) [block, minimum height=10pt, name=server7]{Server 1};
    \draw node at (-3, -6.1) [block, minimum height=10pt, name=server8]{Server 2};
    \draw node at (-3, -6.7) [block, minimum height=10pt, name=server9]{Server $n$};
    \draw node at (0, -6.1) [block, name=request3]{Extract};

    \draw node at (-0.2, -5.9) [block]{};
    \draw node at (-0.1, -6.0) [block]{};
    \draw node at (-0.72, -5.5) [name=request3-1]{};           
    \draw node at (0, -6.1) [block, name=request3-2]{Extract}; 
    \draw node at (-0.72, -6.7) [name=request3-3]{};           

    \draw node at (2.8, -5.9) [block]{};
    \draw node at (2.9, -6.0) [block]{};
    \draw node at (3, -6.1) [block, name=subset3, label={[yshift=12pt]center:Subset}]{};
    \draw node at (3, -6.43) [block, minimum height=20pt, minimum width=40pt, name=buffer3]{Buffer};

    \draw[dashed, ->] ([yshift=-0.3cm]subset3.east) -| ([yshift=-0.3cm]users.south);  

    \draw node at (5.8, -5.9) [block]{};
    \draw node at (5.9, -6.0) [block]{};
    \draw node at (6, -6.1) [block, name=write3]{Write};
    
    \draw node at (8.8, -2.8) [block]{};
    \draw node at (8.9, -2.9) [block]{};
    \draw node at (9, -3) [block, name=store, label={[above=0.25]Store}]{};
    \draw node at (9, -2.67) [block, minimum height=20pt, minimum width=40pt, name=agency]{Agency};
    \draw node at (9, -3.33) [block, minimum height=20pt, minimum width=40pt, name=events]{Events};

    \draw[->] (server1) -- (request1-1);
    \draw[->] (server2) -- (request1-2);
    \draw[->] (server3) -- (request1-3);
    \draw[->] (request) -- (subset);
    \draw[->] ([yshift=-0.3cm]subset.east) -- ([yshift=-0.3cm]write.west);
    \draw[->] (write.east) -- (store);

    \draw[->] (server4) -- (request2-1);
    \draw[->] (server5) -- (request2-2);
    \draw[->] (server6) -- (request2-3);
    \draw[->] (request2-2) -- (subset2);
    \draw[->] ([yshift=-0.3cm]subset2.east) -- ([yshift=-0.3cm]write2.west);
    \draw[->] ([yshift=-0.3cm]write2.east) -- ([yshift=-0.3cm]store.west);

    \draw[->] (server7) -- (request3-1);
    \draw[->] (server8) -- (request3-2);
    \draw[->] (server9) -- (request3-3);
    \draw[->] (request3) -- (subset3);
    \draw[->] ([yshift=0.3cm]subset3.east) -- ([yshift=0.3cm]write3.west);
    \draw[->] (write3.east) -- (store);
    \draw[->] ([yshift=-0.3cm]store.east) -- ([yshift=-0.3cm]users.west);

    \draw node at (-3, 1.45) [name=region1]{\textbf{Region 1}};
    \draw[dotted] (-4, 1.21) rectangle (7, -1);
    
    \draw node at (-3, -1.55) [name=region2]{\textbf{Region 2}};
    \draw[dotted] (-4, -1.79) rectangle (7, -4);
    
    \draw node at (-3, -4.65) [name=region3]{\textbf{Region $N$}};
    \draw[dotted] (-4, -4.89) rectangle (7, -7.1);

    \foreach \x in {1.4, 4.4} { 
        \foreach \y in {-4.25, -4.45, -4.65} { 
            \draw[fill=black] (\x,\y) circle (0.5pt);
        }
    }
\end{tikzpicture}
\caption{ETL Microservice System Design of IDEF0 Architecture}
\label{fig:sys-des}
\end{figure}

\subsection{Validation}
\label{sub-sec:tech}
Realtime data quality was assessed using three separate 24-hour validation periods. As previously mentioned in \nameref{sub-sec:stc-fil}, the first was conducted from December 31, 2023 00:01:30 UTC to January 1, 2024 00:01:30 UTC using \verb|test-data-a.csv|. The second from January 4, 2024 01:30:30 UTC to January 5, 2024 01:30:30 UTC using \verb|test-data-b.csv|. The third from January 10, 2024 16:05:30 UTC to January 11, 2024 16:05:30 UTC using \verb|test-data-c.csv|.\\

All three validation periods were arbitrarily specified and enforced a 72-hour minimum resting period to proxy independence. A battery of 113 individual tests were conducted for each 24-hour validation period to ensure realtime data quality throughout six overlapping categories: \nameref{sub-sec:dat-typ}, \nameref{sub-sec:dup}, \nameref{sub-sec:mis}, \nameref{sub-sec:spl-pnt}, \nameref{sub-sec:tmp-cnt}, and \nameref{sub-sec:dur-exp}. Each category and its corresponding tests are explained in greater detail below.

\subsubsection{Data Types}
\label{sub-sec:dat-typ}

All data type test results were as expected for all validation periods. Exhibited in Table \ref{tab:dat-typ}, test number 1 found the expected data object. Test numbers 2 to 7 found obvious \texttt{string} types - tests 8 to 10, also \texttt{string} types. Note that \textbf{vehicle\_id}, \textbf{route\_id}, and \textbf{trip\_id} contained numerical values in some cases, but were found to be correctly encoded as \texttt{string} types. Geocoordinates from test numbers 12 and 13 were correctly encoded as \texttt{float} types. Test number 13 for \textbf{datetime} was correctly encoded as an \texttt{integer} type, as well as \textbf{duration} as an \texttt{integer} type in test number 14.

\begin{table}[H]
\begin{center}
    \resizebox{\textwidth}{!}{
        \begin{tabular}{|m{0.04\textwidth}|m{0.28\textwidth}|m{0.18\textwidth}|m{0.18\textwidth}|m{0.18\textwidth}|}
        \hline
        \textbf{No.} & \textbf{Field} & \texttt{test-data-a.csv} & \texttt{test-data-b.csv} & \texttt{test-data-c.csv}\\
        \hline
        1 & Object & \texttt{DataFrame} & \texttt{DataFrame} & \texttt{DataFrame} \\
        \hline
        2 & \textbf{iata\_id} & \texttt{string} & \texttt{string} & \texttt{string} \\
        \hline
        3 & \textbf{agency} & \texttt{string} & \texttt{string} & \texttt{string} \\
        \hline
        4 & \textbf{city} & \texttt{string} & \texttt{string} & \texttt{string} \\
        \hline
        5 & \textbf{country} & \texttt{string} & \texttt{string} & \texttt{string} \\
        \hline
        6 & \textbf{region} & \texttt{string} & \texttt{string} & \texttt{string} \\
        \hline
        7 & \textbf{continent} & \texttt{string} & \texttt{string} & \texttt{string} \\
        \hline
        8 & \textbf{vehicle\_id} & \texttt{string} & \texttt{string} & \texttt{string} \\
        \hline
        9 & \textbf{route\_id} & \texttt{string} & \texttt{string} & \texttt{string} \\
        \hline
        10 & \textbf{trip\_id} & \texttt{string} & \texttt{string} & \texttt{string} \\
        \hline
        11 & \textbf{latitude} & \texttt{float} & \texttt{float} & \texttt{float} \\
        \hline
        12 & \textbf{longitude} & \texttt{float} & \texttt{float} & \texttt{float} \\
        \hline
        13 & \textbf{datetime} & \texttt{integer} & \texttt{integer} & \texttt{integer} \\
        \hline
        14 & \textbf{duration} & \texttt{integer} & \texttt{integer} & \texttt{integer} \\
        \hline
        \end{tabular}
    }
\caption{Global Data Types by Field}
\label{tab:dat-typ}
\end{center}
\end{table}

\subsubsection{Duplication}
\label{sub-sec:dup}

Of the 192,237 observations in \texttt{test-data-a.csv}, 197,411 observations in \texttt{test-data-b.csv}, and 209,329 observations in \texttt{test-data-c.csv}, none of them contained duplicate observations, nor did they contain duplicate fields. Although \nameref{sub-sec:dat-typ} previously implied that all validation periods contained the same unique fields, Table \ref{tab:dup} exhibits this explicitly from the results of test number 15. In addition to all unique fields, test number 16 found all unique observations.

\begin{table}[H]
\begin{center}
    \resizebox{\textwidth}{!}{
        \begin{tabular}{|m{0.04\textwidth}|m{0.28\textwidth}|>{\centering\arraybackslash}m{0.18\textwidth}|>{\centering\arraybackslash}m{0.18\textwidth}|>{\centering\arraybackslash}m{0.18\textwidth}|}
        \hline
        \textbf{No.} & \textbf{Dimension} & \texttt{test-data-a.csv} & \texttt{test-data-b.csv} & \texttt{test-data-c.csv}\\
        \hline
        15 & Fields & 0.00 \% & 0.00 \% & 0.00 \% \\
        \hline
        16 & Observations & 0.00 \% & 0.00 \% & 0.00 \% \\
        \hline
        \end{tabular}
}
\caption{Global Percentages of Duplication by Dimension}
\label{tab:dup}
\end{center}
\end{table}

\subsubsection{Missingness}
\label{sub-sec:mis}

 Two of thirteen fields contained missing values across all validation periods. In Table \ref{tab:mis}, test number 24 found that \textbf{route\_id} averaged (unweighted) 7.66\% missingness. Test number 25 found that \textbf{trip\_id} averaged (unweighted) 1.38\% missingness. Although both \textbf{route\_id} and \textbf{trip\_id} contained missing values, when tested together, no missing values were present. Test number 26 found that every observation contained either a \textbf{route\_id} or \textbf{trip\_id} or both.

\begin{table}[H]
\begin{center}
    \resizebox{\textwidth}{!}{
        \begin{tabular}{|m{0.04\textwidth}|m{0.28\textwidth}|>{\centering\arraybackslash}m{0.18\textwidth}|>{\centering\arraybackslash}m{0.18\textwidth}|>{\centering\arraybackslash}m{0.18\textwidth}|}
        \hline
        \textbf{No.} & \textbf{Field} & \texttt{test-data-a.csv} & \texttt{test-data-b.csv} & \texttt{test-data-c.csv}\\
        \hline
        17 & \textbf{iata\_id} & 0.00 \% & 0.00 \% & 0.00 \% \\
        \hline
        18 & \textbf{agency} & 0.00 \% & 0.00 \% & 0.00 \% \\
        \hline
        19 & \textbf{city} & 0.00 \% & 0.00 \% & 0.00 \% \\
        \hline
        20 & \textbf{country} & 0.00 \% & 0.00 \% & 0.00 \% \\
        \hline
        21 & \textbf{region} & 0.00 \% & 0.00 \% & 0.00 \% \\
        \hline
        22 & \textbf{continent} & 0.00 \% & 0.00 \% & 0.00 \% \\
        \hline
        23 & \textbf{vehicle\_id} & 0.00 \% & 0.00 \% & 0.00 \% \\
        \hline
        24 & \textbf{route\_id} & 7.50 \% & 8.67 \% & 6.82 \% \\
        \hline
        25 & \textbf{trip\_id} & 1.64 \% & 1.31 \% & 1.20 \% \\
        \hline
        26 & \textbf{route\_id | trip\_id} & 0.00 \% & 0.00 \% & 0.00 \% \\
        \hline
        27 & \textbf{latitude} & 0.00 \% & 0.00 \% & 0.00 \% \\
        \hline
        28 & \textbf{longitude} & 0.00 \% & 0.00 \% & 0.00 \% \\
        \hline
        29 & \textbf{datetime} & 0.00 \% & 0.00 \% & 0.00 \% \\
        \hline
        30 & \textbf{duration} & 0.00 \% & 0.00 \% & 0.00 \% \\
        \hline
        \end{tabular}
    }
\caption{Global Percentages of Missingness by Field(s)}
\label{tab:mis}
\end{center}
\end{table}

\subsubsection{Geolocation Error}
\label{sub-sec:spl-pnt}

General geolocation tests did not find any value errors across all validation periods. Table \ref{tab:spl-pnt-lat} and Table \ref{tab:spl-pnt-lon} exhibit the test results in the \textbf{latitude} and \textbf{longitude} fields, respectively. Tests 31 to 33 corresponding to \textbf{latitude} and tests 34 to 36 corresponding to \textbf{longitude}, did not find any zero values or those outside the projected bounds of the World Geodetic System 1984 (WGS84). Recall tests 11 and 12 in \nameref{sub-sec:dat-typ}, and 27 and 28 in \nameref{sub-sec:mis}, also did not contain value errors.

\begin{table}[H]
\begin{center}
    \resizebox{\textwidth}{!}{
        \begin{tabular}{|m{0.04\textwidth}|m{0.28\textwidth}|>{\centering\arraybackslash}m{0.18\textwidth}|>{\centering\arraybackslash}m{0.18\textwidth}|>{\centering\arraybackslash}m{0.18\textwidth}|}
        \hline
        \textbf{No.} & \textbf{latitude} & \texttt{test-data-a.csv} & \texttt{test-data-b.csv} & \texttt{test-data-c.csv}\\
        \hline
        31 & Zero Values & 0.00 \% & 0.00 \% & 0.00 \% \\
        \hline
        32 & Exceed Max. ($> \text{\space} 90^{\circ}$) & 0.00 \% & 0.00 \% & 0.00 \% \\
        \hline
        33 & Exceed Min. ($<-90^{\circ}$)& 0.00 \% & 0.00 \% & 0.00 \% \\
        \hline
        \end{tabular}
}
\caption{Global Percentages of Erroneous Values in \textbf{latitude} Field (WGS84)}
\label{tab:spl-pnt-lat}
\end{center}
\end{table}

\begin{table}[H]
\begin{center}
    \resizebox{\textwidth}{!}{
        \begin{tabular}{|m{0.04\textwidth}|m{0.28\textwidth}|>{\centering\arraybackslash}m{0.18\textwidth}|>{\centering\arraybackslash}m{0.18\textwidth}|>{\centering\arraybackslash}m{0.18\textwidth}|}
        \hline
        \textbf{No.} & \textbf{longitude} & \texttt{test-data-a.csv} & \texttt{test-data-b.csv} & \texttt{test-data-c.csv}\\
        \hline
        34 & Zero Values & 0.00 \% & 0.00 \% & 0.00 \% \\
        \hline
        35 & Exceed Max. ($> \text{\space} 180^{\circ}$) & 0.00 \% & 0.00 \% & 0.00 \% \\
        \hline
        36 & Exceed Min. ($<-180^{\circ}$)& 0.00 \% & 0.00 \% & 0.00 \% \\
        \hline
        \end{tabular}
}
\caption{Global Percentages of Erroneous Values in \textbf{longitude} Field (WGS84)}
\label{tab:spl-pnt-lon}
\end{center}
\end{table}

In addition to the value error tests previously mentioned, each idling event was tested to ensure the recorded geolocation was within a reasonable proximity to the shape of the corresponding route path. The shape of the route paths and their identifiers were collected from the GTFS static data corresponding to the GTFS Realtime source. The shape of the route paths were used as a geographic benchmark to validate the measured geolocation of a given idling event - anticipating the \textbf{route\_id} or \textbf{trip\_id} should be geographically near the shape of its corresponding route path. Recall test 26 from \nameref{sub-sec:mis} found that every observation contained either a \textbf{route\_id}, \textbf{trip\_id}, or both.\\
 
The notion of reasonable proximity was operationalized as a distance threshold specified with three primary considerations. First, the width of a typical urban arterial street was used as a tolerable margin of error between approximately 15 to 20 meters throughout varying degrees of urban density \cite{aashto2018, nacto2012, southworth1995}. Second, Global Positioning System (GPS) location errors were estimated between 5 to 10 meters, especially in the context of urban environments where buildings, infrastructure, etc. can all obstruct reported vehicle locations \cite{merry2019, beekhuizen2013, vandiggelen2015}. Third, a preliminary test was drawn on a series of distance thresholds that combined the first two considerations and specified its midpoint to ensure that 25 meters was flexible enough to afford meaningful results, as exhibited in Figure \ref{fig:spl-pnt-dst}.\\

\begin{tikzpicture}
\begin{axis}[
    title={Figure \ref{fig:spl-pnt-dst} (a): \texttt{test-data-a.csv}},
    xlabel={Distance Threshold $D_m$ (meters)},
    ylabel={Average Spatial Point Error (\%)},
    width=\textwidth,
    height=6cm,
        legend style={
            at={(0.98, 0.95)},
            anchor=north east,
            font=\small,
            fill opacity=0.90,
            text opacity=1,
            legend image post style={yshift=-2pt}
        },  
    ymajorgrids=true,
    xmin=-1, xmax=101,
    xtick={0,10,...,100},
    ytick={0,10,...,100},
    xtick pos=left,
    ytick pos=left,
    grid style=dashed,
    extra x ticks={25},
    extra x tick labels={25},
    extra x tick style={grid=major},
]

\draw[dashed, line width=.75pt, color=gray] (axis cs:25,\pgfkeysvalueof{/pgfplots/ymin}) -- (axis cs:25,\pgfkeysvalueof{/pgfplots/ymax});

\addplot[
    color=yellow!60!black,
    dashed,
    mark=none,
    line width=1.5pt
    ]
    coordinates {
    (0,96.92467384776477)(1,19.156486303481927)(2,12.94528348424636)(3,11.073901842363938)(4,9.932107554178302)(5,9.23700321973773)(6,8.686814730276879)(7,8.187880080028222)(8,7.926956928826783)(9,7.736706445656196)(10,7.522058878206636)(11,7.354839323120714)(12,7.2456415019129)(13,7.062586892250565)(14,6.919000616108832)(15,6.7660364436526805)(16,6.612874395819958)(17,6.466065198040127)(18,6.341988177808574)(19,6.212903648894283)(20,6.148256895119502)(21,6.090825802837713)(22,5.9242475468583535)(23,5.838234700594384)(24,5.795196401096845)(25,5.72715970329854)(26,5.641102127104034)(27,5.500659382850884)(28,5.377089252581925)(29,5.299834198281772)(30,5.228559196485634)(31,5.161344220306091)(32,5.068119544852792)(33,5.031284787918494)(34,4.948234607289351)(35,4.9149636621786925)(36,4.880634517445756)(37,4.843637643715425)(38,4.762322999117842)(39,4.72008917026686)(40,4.668399809330822)(41,4.465452172922156)(42,4.415839626926882)(43,4.276780965690008)(44,4.162839062933074)(45,4.110356593983539)(46,4.041880264906922)(47,3.948487852827924)(48,3.812646810135206)(49,3.7150263836612822)(50,3.688716978283722)(51,3.605412131159852)(52,3.5479082195426486)(53,3.5100857253299296)(54,3.483887006420673)(55,3.431699238470401)(56,3.4058413648193806)(57,3.367032943337847)(58,3.3477424852548268)(59,3.317868622512691)(60,3.2688751530404687)(61,3.2260426884822095)(62,3.1868240699007373)(63,3.160188540346411)(64,3.141504971874369)(65,3.0993204075196985)(66,3.0668241607015716)(67,3.051183337211242)(68,3.006125380837503)(69,2.933672141322049)(70,2.9167974419318057)(71,2.9011127247236317)(72,2.868610384859821)(73,2.8246901542969596)(74,2.8028768024317827)(75,2.779705256016129)(76,2.7430476276400384)(77,2.719994385256527)(78,2.6926701164761226)(79,2.6733381020747515)(80,2.657236726081564)(81,2.6428851125689903)(82,2.6203029735985126)(83,2.6054538067129585)(84,2.5867401526595586)(85,2.571656538676365)(86,2.5628165225066795)(87,2.5467781312245137)(88,2.517265294050773)(89,2.484357266512248)(90,2.456621108220304)(91,2.4377374068420608)(92,2.4053241956486033)(93,2.394114479140441)(94,2.3583819342013044)(95,2.332074604618896)(96,2.3105478167145606)(97,2.2975199490079916)(98,2.2823938402840582)(99,2.258452639717092)(100,2.132398439047904)
    };

\addplot[
    color=yellow!60!black,
    solid,
    mark=none,
    line width=1pt
    ]
    coordinates {
    (0,89.943730343463)(1,18.669440564926802)(2,11.889141838936212)(3,10.147227870357582)(4,9.214585309126079)(5,8.707645903835967)(6,8.428954702433671)(7,8.21536039450136)(8,8.004039019222674)(9,7.815744829484089)(10,7.51770941159414)(11,7.374319064595909)(12,7.268492189529724)(13,7.1161226080212)(14,6.909396954277938)(15,6.612947987132131)(16,6.341887345352859)(17,6.203688437897142)(18,6.021849990510077)(19,5.83679006377136)(20,5.708066396376864)(21,5.554204963121023)(22,5.350393981172047)(23,5.235217145797165)(24,5.165707336961404)(25,5.05471331128436)(26,4.966447333779072)(27,4.8170461153874555)(28,4.7206369539654816)(29,4.617807844407025)(30,4.537685050818382)(31,4.455890156042443)(32,4.36894345103353)(33,4.306694754387166)(34,4.240653149313772)(35,4.208791184488234)(36,4.174984826768338)(37,4.136417838451052)(38,4.0917262042711116)(39,4.056420324614919)(40,3.9974314582648987)(41,3.8516505690493688)(42,3.7922973687250874)(43,3.752867484152844)(44,3.6989844069021416)(45,3.6689747927038496)(46,3.6388731362383595)(47,3.5368455614233483)(48,3.49456108820344)(49,3.436234414831233)(50,3.376684346324282)(51,3.2397880926223053)(52,3.194659257673237)(53,3.1703281400145897)(54,3.1403530416664864)(55,3.1002430670089325)(56,3.067764930339152)(57,3.0366968300681663)(58,3.0114746921285445)(59,2.975617837234026)(60,2.9375647515751666)(61,2.909040597300134)(62,2.883479692682627)(63,2.8679398899054576)(64,2.84769953967745)(65,2.8122453697019125)(66,2.7742204080691693)(67,2.7449957098715885)(68,2.6672468397575386)(69,2.6431982407245727)(70,2.629100433466661)(71,2.609179674444192)(72,2.4893495911168912)(73,2.448827982986657)(74,2.4170005340113505)(75,2.3930861632846785)(76,2.366925705565663)(77,2.33237022606005)(78,2.28052614070063)(79,2.241755892376588)(80,2.230552303075882)(81,2.216489011668159)(82,2.195674675635061)(83,2.178107386250076)(84,2.157419608334351)(85,2.144506845266477)(86,2.132266502095078)(87,2.1138158626180257)(88,2.0867643846180073)(89,2.0613581621440034)(90,2.037777444635921)(91,2.017779983724111)(92,1.9808863749588852)(93,1.9687802600938085)(94,1.9480323990314616)(95,1.9238994279202473)(96,1.9072960256110463)(97,1.8958840628443312)(98,1.8816251816188583)(99,1.8561524841741175)(100,1.8412390801597809)
    };
        \legend{
            Unweighted,
            Weighted
        }
\end{axis}
\end{tikzpicture}

\begin{tikzpicture}
\begin{axis}[
    title={Figure \ref{fig:spl-pnt-dst} (b): \texttt{test-data-b.csv}},
    xlabel={Distance Threshold $D_m$ (meters)},
    ylabel={Average Spatial Point Error (\%)},
    width=\textwidth,
    height=6cm,
        legend style={
            at={(0.98, 0.95)},
            anchor=north east,
            font=\small,
            fill opacity=0.90,
            text opacity=1,
            legend image post style={yshift=-2pt}
        },  
    ymajorgrids=true,
    xmin=-1, xmax=101,
    xtick={0,10,...,100},
    ytick={0,10,...,100},
    xtick pos=left,
    ytick pos=left,
    grid style=dashed,
    extra x ticks={25},
    extra x tick labels={25},
    extra x tick style={grid=major},
]

\draw[dashed, line width=.75pt, color=gray] (axis cs:25,\pgfkeysvalueof{/pgfplots/ymin}) -- (axis cs:25,\pgfkeysvalueof{/pgfplots/ymax});

\addplot[
    color=orange!60!black,
    dashed,
    mark=none,
    line width=1.5pt
    ]
    coordinates {
    (0,97.02232503830336)(1,20.64752286125278)(2,14.261986676726181)(3,11.666182126188247)(4,10.561496878430276)(5,9.641466302917749)(6,9.256981476771458)(7,8.904189661170278)(8,8.715603443262637)(9,8.474965754209322)(10,8.260256040613271)(11,8.06150148641909)(12,7.920730159489693)(13,7.748626758629598)(14,7.598915458296986)(15,7.4803712857587215)(16,7.3569855339669346)(17,7.221722045320121)(18,7.0768894251091155)(19,6.958524316010397)(20,6.8834708775477935)(21,6.82725414563501)(22,6.719619613397512)(23,6.6523814992746395)(24,6.584441887925365)(25,6.502597539657657)(26,6.38894830702975)(27,6.286511906334283)(28,6.196930221864349)(29,6.076751655999445)(30,5.98839378637193)(31,5.917985980767014)(32,5.835043897487495)(33,5.736394090292592)(34,5.681414275467162)(35,5.642444114816314)(36,5.614628621460255)(37,5.565549208382905)(38,5.4702546827023895)(39,5.4184731159448205)(40,5.374744804719072)(41,5.24471192044831)(42,5.1949088947294655)(43,5.07976102952216)(44,4.963333796563077)(45,4.914354701862052)(46,4.865799736056076)(47,4.796258940810105)(48,4.698298106564252)(49,4.569330643277951)(50,4.531178835165832)(51,4.454092738746738)(52,4.416557259153919)(53,4.390299714255889)(54,4.347132210764187)(55,4.296313579644234)(56,4.246246120571598)(57,4.1970079387436385)(58,4.164224610298845)(59,4.110491757759476)(60,4.078623521426479)(61,4.036031082830604)(62,4.005300823195526)(63,3.9738375606154506)(64,3.9329848047686085)(65,3.880258753457511)(66,3.853735535283306)(67,3.822278706256469)(68,3.772206580322333)(69,3.720646718625389)(70,3.670185132875176)(71,3.622507128776121)(72,3.5727098934093107)(73,3.5322090196234655)(74,3.421711369544809)(75,3.389602434337476)(76,3.363922173661365)(77,3.345347248364405)(78,3.315010327340829)(79,3.293670457696095)(80,3.2670273531981735)(81,3.246573453896005)(82,3.2111518688290346)(83,3.18943653242178)(84,3.1644460147394824)(85,3.1468218928329748)(86,3.132861382169253)(87,3.1039539189508076)(88,3.0789276436891697)(89,3.050433643103574)(90,3.0298397673211213)(91,3.0132071412062515)(92,2.977381587653184)(93,2.9591667564227038)(94,2.9421719905849244)(95,2.9211565563716277)(96,2.896164659392113)(97,2.8854594478175244)(98,2.8731436625854627)(99,2.849064348608721)(100,2.7790380042777514)
    };
    \addlegendentry{Test B - Unweighted}

\addplot[
    color=orange!60!black,
    solid,
    line width=1pt,
    mark=none
    ]
    coordinates {
    (0,88.85568010650384)(1,18.5605138226112)(2,11.639597833557687)(3,9.73826087543597)(4,8.708316080896026)(5,8.225943137337993)(6,7.941398006764828)(7,7.6825539764848685)(8,7.474693514802453)(9,7.2533497153145134)(10,6.986039881496637)(11,6.803086813215785)(12,6.699281808995465)(13,6.522280332482822)(14,6.354514168018952)(15,6.102507390246046)(16,5.921511970975317)(17,5.785864317403991)(18,5.6149572031066555)(19,5.4547542425750635)(20,5.308767830788696)(21,5.15463261425343)(22,4.96541942484852)(23,4.86767611502934)(24,4.768490581711944)(25,4.671651081417181)(26,4.598903489089551)(27,4.467451834756346)(28,4.359513747088073)(29,4.27925497734681)(30,4.2109066506563835)(31,4.129567423210474)(32,4.056967441005966)(33,3.981819369298748)(34,3.939991258092931)(35,3.9095751032926245)(36,3.8873948674121124)(37,3.850148719618929)(38,3.8030756093654503)(39,3.7618137402713927)(40,3.714600279409069)(41,3.573702787113831)(42,3.5284445554236044)(43,3.4798384777437406)(44,3.439738130930124)(45,3.4076123605232596)(46,3.3719112793468753)(47,3.3102320264167275)(48,3.264241793705949)(49,3.196770658321114)(50,3.133306774803799)(51,2.986064467051591)(52,2.9432720503692025)(53,2.920144447878755)(54,2.877529889295559)(55,2.836261970606273)(56,2.788777487878491)(57,2.7644133481269364)(58,2.7372010589501485)(59,2.7030704526904685)(60,2.670285276405636)(61,2.6366156493009782)(62,2.600492306378868)(63,2.5698474768795876)(64,2.5426121992410344)(65,2.508208151277813)(66,2.46997107978909)(67,2.433283914593389)(68,2.413098835213347)(69,2.3931944570511234)(70,2.3740390187759743)(71,2.34799309155585)(72,2.240597047618465)(73,2.2000127331879895)(74,2.1608367645225712)(75,2.140550051942995)(76,2.11407944313909)(77,2.084450945635197)(78,2.044338499631152)(79,2.0144147818809586)(80,1.9946410749753056)(81,1.977022233883659)(82,1.954395537876664)(83,1.9390368255557746)(84,1.9200882834367974)(85,1.900504533820694)(86,1.8891530733673818)(87,1.8687647275882853)(88,1.8425869191925557)(89,1.8212330580178673)(90,1.8051665430572825)(91,1.7852428061903396)(92,1.757567118020475)(93,1.7412695085229046)(94,1.7202943520083664)(95,1.6931498182979965)(96,1.6780185707570183)(97,1.6690518607356921)(98,1.6570143761854723)(99,1.6394620806411666)(100,1.6207458429783372)
    };
        \legend{
            Unweighted,
            Weighted
        }
\end{axis}
\end{tikzpicture}

\begin{figure}[!ht]
\centering

\begin{tikzpicture}
\begin{axis}[
    title={Figure \ref{fig:spl-pnt-dst} (c): \texttt{test-data-c.csv}},
    xlabel={Distance Threshold $D_m$ (meters)},
    ylabel={Average Spatial Point Error (\%)},
    width=\textwidth,
    height=6cm,
        legend style={
            at={(0.98, 0.95)},
            anchor=north east,
            font=\small,
            fill opacity=0.90,
            text opacity=1,
            legend image post style={yshift=-2pt}
        },  
    ymajorgrids=true,
    xmin=-1, xmax=101,
    xtick={0,10,...,100},
    ytick={0,10,...,100},
    xtick pos=left,
    ytick pos=left,
    grid style=dashed,
    extra x ticks={25},
    extra x tick labels={25},
    extra x tick style={grid=major},
]

\draw[dashed, line width=.75pt, color=gray] (axis cs:25,\pgfkeysvalueof{/pgfplots/ymin}) -- (axis cs:25,\pgfkeysvalueof{/pgfplots/ymax});

\addplot[
    color=blue!80!black,
    dashed,
    mark=none,
    line width=1.5pt
    ]
    coordinates {
    (0,97.00901728085205)(1,20.486123032333182)(2,13.931877353636821)(3,11.201236468846886)(4,10.139008393886584)(5,9.358785111716514)(6,8.900240889875278)(7,8.551597335261931)(8,8.32503888303971)(9,8.128823050882545)(10,7.929293456005922)(11,7.786456598621442)(12,7.678868426802765)(13,7.403248064554262)(14,7.17800920101125)(15,7.019231575469377)(16,6.887823064827629)(17,6.742554746805794)(18,6.626310676585774)(19,6.458889377408468)(20,6.3978443732723065)(21,6.309396483036707)(22,6.199666557128253)(23,6.116263262974812)(24,6.053391262379094)(25,5.957151862473879)(26,5.800572337643786)(27,5.640857710004724)(28,5.556007275176626)(29,5.4657959886256435)(30,5.370291064361709)(31,5.178605766368804)(32,5.042381051174914)(33,4.92650361298584)(34,4.848822112594689)(35,4.8209252710980195)(36,4.7864297544945344)(37,4.750378966899248)(38,4.635907095220773)(39,4.547726017040475)(40,4.49676110469953)(41,4.328870732242095)(42,4.276399656522614)(43,4.117082932776427)(44,3.929473303120133)(45,3.892601195119539)(46,3.8419186405944856)(47,3.7594286555254257)(48,3.7114362745507634)(49,3.629408243577984)(50,3.608982707939191)(51,3.5261797528680603)(52,3.499860235409912)(53,3.448409289262969)(54,3.410350904402776)(55,3.3360323972027004)(56,3.2750776962206913)(57,3.240655708094849)(58,3.2078874037705134)(59,3.164234528940284)(60,3.114689782889343)(61,3.0879786555000948)(62,3.042331160583174)(63,3.0187091781612594)(64,2.99257611716925)(65,2.942296195817562)(66,2.9081179371745804)(67,2.8758756050212497)(68,2.8563549800757357)(69,2.79633424922298)(70,2.750368128028015)(71,2.713196604157062)(72,2.6655729007195106)(73,2.6273979117356703)(74,2.5586573700447417)(75,2.535415959902764)(76,2.5013000036117887)(77,2.469769776851493)(78,2.444632342168674)(79,2.4277279777018634)(80,2.4141234591009066)(81,2.3720028305649947)(82,2.3271012376354463)(83,2.3124682933083562)(84,2.283514876583652)(85,2.2686332677419045)(86,2.2407437351937887)(87,2.211912860080332)(88,2.1922594603341707)(89,2.1652713383765843)(90,2.133851155060924)(91,2.1049486347741464)(92,2.0702261239714517)(93,2.0608609214942533)(94,2.0428986716462845)(95,2.019477623533703)(96,1.9743578192740614)(97,1.9617428163736008)(98,1.9487635538006884)(99,1.9251429299828402)(100,1.9011954465724017)
    };
    \addlegendentry{Test C - Unweighted}

\addplot[
    color=blue!80!black,
    solid,
    mark=none,
    line width=1pt
    ]
    coordinates {
    (0,87.39417263414985)(1,19.539597884992602)(2,12.587199000462022)(3,10.5555868621002)(4,9.449049046895041)(5,8.904692193530053)(6,8.539855435492342)(7,8.269892750054296)(8,8.068936114699426)(9,7.83744192505209)(10,7.551473451116166)(11,7.3658931858055325)(12,7.241552108249252)(13,7.038958715932537)(14,6.8191692886831845)(15,6.54221978942725)(16,6.335775916271203)(17,6.196418904879138)(18,6.010487617738576)(19,5.837724356409566)(20,5.68259826089529)(21,5.538902565888904)(22,5.349502752697049)(23,5.237591344689363)(24,5.122437355716258)(25,4.9974076836122805)(26,4.902755521364924)(27,4.737484760474658)(28,4.63448338933766)(29,4.5562525931569695)(30,4.48967276292511)(31,4.396977446511201)(32,4.303772408742731)(33,4.2065716397224975)(34,4.146637015857166)(35,4.114154719932344)(36,4.077754697606807)(37,4.039475834298898)(38,3.9855153060984634)(39,3.943863405436275)(40,3.899689796225786)(41,3.737692549259066)(42,3.691143826823847)(43,3.6276129103408437)(44,3.5671174584943657)(45,3.537716354187779)(46,3.5020237551670306)(47,3.402621366470214)(48,3.3560027086512463)(49,3.2976483554762126)(50,3.2683978812269174)(51,3.118956725463974)(52,3.0797713919898797)(53,3.0377523305630802)(54,2.9957709266505645)(55,2.952299361099932)(56,2.8559795089325775)(57,2.830896914571909)(58,2.804154699758655)(59,2.772057048444367)(60,2.737324716038799)(61,2.7076828726227973)(62,2.6685688194436494)(63,2.6282120682919157)(64,2.603826137946186)(65,2.573000772651497)(66,2.52993940502472)(67,2.490986741192714)(68,2.457451379778064)(69,2.4383321217924845)(70,2.4186426214471766)(71,2.3965672485873597)(72,2.2838354447319063)(73,2.2425601192129534)(74,2.2066644385970733)(75,2.1872143324499262)(76,2.158733147421785)(77,2.119683649981482)(78,2.080728296326953)(79,2.0502593325071103)(80,2.034334238682746)(81,2.0149244798723913)(82,1.9865441631861671)(83,1.9737190897371875)(84,1.955880187238563)(85,1.9417761032138543)(86,1.930548784301152)(87,1.9105486094627224)(88,1.8911899572789395)(89,1.8689289866699426)(90,1.8546500641858898)(91,1.8308814480928248)(92,1.8031843463092656)(93,1.7927115225941748)(94,1.772662930951583)(95,1.7502325015281457)(96,1.7319726418158012)(97,1.7242837843380983)(98,1.7123638361440499)(99,1.6909367104950377)(100,1.674513999517174)
    };
        \legend{
            Unweighted,
            Weighted
        }
\end{axis}
\end{tikzpicture}

\caption{Global Average Spatial Point Error Relationship to Distance Threshold $D_m$}
\label{fig:spl-pnt-dst}
\end{figure}
\hspace*{\fill}

Spatial point error testing begins with the global distance threshold $D_{m}$ specified in meters:

\begin{equation}
    D_m = 25 \text{m}
\end{equation}
$D_d$ is then computed from $D_m$ by converting meters to degrees, using radians: 

\begin{equation}
D_d = \text{degrees} / \text{m} = 1 / 111.32 \times 1000 \times \cos\left(\phi \times \frac{\pi}{180}\right)
\end{equation}\\
where $\phi$ is the average latitude of a given city in degrees:

\begin{equation}
    \phi = \frac{1}{n}\sum^{n}_{i=1} \text{latitude}_{i}
\end{equation}

A k-Dimensional Tree (K-D Tree) \cite{friedman1977} is constructed for the shape of each route path $s$ in the set of all route path shapes $S$ of a given city, composed of a set of geocoordinates $\{p_1, p_2,\hdots, p_m\}$, where each geocoordinate $p_i$ contains a tuple $(\text{latitude}_i, \text{longitude}_i)$, such that:

\begin{equation}
    \forall \text{\space} s \in S : \text{K-D Tree}_s = KDTree(\{(\text{latitude}_i, \text{longitude}_i) \mid p_i \in s\})    
\end{equation}\\
Given a set of idling events $P$, where each idling event $P_j \in P$ contains a tuple $(\text{latitude}_j, \text{longitude}_j)$, a given $P_j$ within $D_d$ of $s$ is computed as:

\begin{equation}
    C(P_j, D_d) = \sum_{s \in S} \textbf{1} \left( dist(P_j, s) \leq D_d \right)
\end{equation}

where $C(P_j, D_d)$ is a counter of $P_j$ and $D_d$, $\textbf{1}$ is an indicator function, and $dist(P_j, s)$ is the distance between $P_j$ and $s$ using $\text{K-D Tree}_s$.\\

Finally the spatial point error $e$ is computed as the percentage of idling events $P_j$ that are outside the bounds of the global distance threshold $D_d$ from all route path shapes $S$ of a given city:

\begin{equation}
    e = 1 - \left( \frac{\sum_{j=1}^{n} C(P_j, D_d)}{n} \right)
\end{equation}\\
If $e$ is 0\%, all idling events were within the distance threshold corresponding to the shape of the route path. Any $e$ greater than 0\% indicates that there were idling events measured outside of it.\\

Table \ref{tab:spl-pnt} exhibits the global average spatial point errors for all validation periods. The reported averages are both weighted and unweighted. Weighted averages are weighted by the number of observations per city, whereas the unweighted averages assign equal weight to all cities within the context of how they are reported, globally or regionally. Table \ref{tab:spl-pnt-us-east} to Table \ref{tab:spl-pnt-asia} exhibit the test results for individual cities and their regional unweighted and weighted averages.\\

Globally, test numbers 37 and 38 found the unweighted and weighted average error did not exceed 6.50\% and 5.05\%, respectively. That is to say, the global average spatial point accuracy of measured idling events was at least 93.50\%. Regionally, test number 52 in Table \ref{tab:spl-pnt-us-west} found that 11.06\% was the highest unweighted average error. Test number 67 in Table \ref{tab:spl-pnt-us-cent} found that 12.29\% was the highest weighted average error. In other words, the average spatial point accuracy of measured idling events for any given region was at least 87.71\%.\\

Note that in Table \ref{tab:spl-pnt-eu-west}, Table \ref{tab:spl-pnt-eu-cent}, and Table \ref{tab:spl-pnt-asia}, cities without results encountered preliminary test errors. In those cases, identity mappings could not be established between \textbf{route\_id} or \textbf{trip\_id} and the shape of the corresponding route path, either because of unmatched identifiers or the shape of the corresponding route path was unavailable. In successful preliminary tests, the city is indicated along with the GTFS static data source and the geolocation error.

\begin{table}[H]
\begin{center}
    \resizebox{\textwidth}{!}{
        \begin{tabular}{|m{0.04\textwidth}|m{0.28\textwidth}|>{\centering\arraybackslash}m{0.18\textwidth}|>{\centering\arraybackslash}m{0.18\textwidth}|>{\centering\arraybackslash}m{0.18\textwidth}|}
        \hline
        \textbf{No.} & \textbf{Global} & \texttt{test-data-a.csv} & \texttt{test-data-b.csv} & \texttt{test-data-c.csv}\\
        \hline
        37 & \textbf{Average (Unweighted)} & \textbf{5.73 \%} & \textbf{6.50 \%} & \textbf{5.96 \%} \\
        \hline
        38 & \textbf{Average (Weighted)} & \textbf{5.05 \%} & \textbf{4.67 \%} & \textbf{5.00 \%} \\
        \hline
        \end{tabular}
    }
\caption{Global Percentages of \textbf{latitude -- longitude} 25m Outside GTFS Routes (WGS84)}
\label{tab:spl-pnt}
\end{center}
\end{table}

\begin{table}[H]
\begin{center}
    \resizebox{\textwidth}{!}{
        \begin{tabular}{|m{0.04\textwidth}|m{0.14\textwidth}|m{0.12\textwidth}|>{\centering\arraybackslash}m{0.18\textwidth}|>{\centering\arraybackslash}m{0.18\textwidth}|>{\centering\arraybackslash}m{0.18\textwidth}|}
        \hline
        \textbf{No.} & \textbf{City} & \textbf{Routes} & \texttt{test-data-a.csv} & \texttt{test-data-b.csv} & \texttt{test-data-c.csv}\\
        \hline
        39 & New York & GTFS \cite{mta-st} & 2.93 \% & 2.97 \% & 2.87 \% \\
        \hline
        40 & Philadelphia & GTFS \cite{septa-st} & 6.72 \% & 8.16 \% & 6.39 \% \\
        \hline
        41 & \textit{Wash. D.C.} & GTFS \cite{wmata-st} & 5.24 \% & 5.60 \% & 5.97 \% \\
        \hline
        42 & Boston & GTFS \cite{mbta-st} & 9.78 \% & 7.84 \% & 7.31 \% \\
        \hline
        43 & Pittsburgh & GTFS \cite{prt-st} & 2.49 \% & 7.95 \% & 6.91 \% \\
        \hline
        44 & \multicolumn{2}{|l|}{\textbf{Average (Unweighted)}} &  \textbf{5.43 \%} & \textbf{6.50 \%} & \textbf{5.89 \%} \\
        \hline
        45 & \multicolumn{2}{|l|}{\textbf{Average (Weighted)}} & \textbf{4.24 \%} & \textbf{3.95 \%} & \textbf{3.79 \%} \\
        \hline
        \end{tabular}
    }
\caption{US East Percentages of \textbf{latitude -- longitude} 25m Outside GTFS Routes (WGS84)}
\label{tab:spl-pnt-us-east}
\end{center}
\end{table}

\begin{table}[H]
\begin{center}
    \resizebox{\textwidth}{!}{
        \begin{tabular}{|m{0.04\textwidth}|m{0.14\textwidth}|m{0.12\textwidth}|>{\centering\arraybackslash}m{0.18\textwidth}|>{\centering\arraybackslash}m{0.18\textwidth}|>{\centering\arraybackslash}m{0.18\textwidth}|}
        \hline
        \textbf{No.} & \textbf{City} & \textbf{Routes} & \texttt{test-data-a.csv} & \texttt{test-data-b.csv} & \texttt{test-data-c.csv}\\
        \hline
        46 & Los Angeles & GTFS \cite{la-metro-st} & 11.75 \% & 11.36 \% & 12.76 \% \\
        \hline
        47 & \textit{San Fran.} & GTFS \cite{mtc-st} & 5.29 \% & 5.13 \% & 5.26 \% \\
        \hline
        48 & San Diego & GTFS \cite{mts-st} & 3.91 \% & 5.99 \% & 6.79 \% \\
        \hline
        49 & Seattle & GTFS \cite{kc-metro-st} & 2.10 \% & 2.49 \% & 1.91 \% \\
        \hline
        50 & Sacrameto & GTFS \cite{sacrt-st} & 6.71 \% & 35.04 \% & 5.11 \% \\
        \hline
        51 & Portland & GTFS \cite{trimet-st} & 8.11 \% & 6.35 \% & 8.80 \% \\
        \hline
        52 & \multicolumn{2}{|l|}{\textbf{Average (Unweighted)}} &  \textbf{6.31 \%} & \textbf{11.06 \%} & \textbf{6.77 \%} \\
        \hline
        53 & \multicolumn{2}{|l|}{\textbf{Average (Weighted)}} & \textbf{6.17 \%} & \textbf{5.72 \%} & \textbf{6.27 \%} \\
        \hline
        \end{tabular}
    }
\caption{US West Percentages of \textbf{latitude -- longitude} 25m Outside GTFS Routes (WGS84)}
\label{tab:spl-pnt-us-west}
\end{center}
\end{table}

\begin{table}[H]
\begin{center}
    \resizebox{\textwidth}{!}{
        \begin{tabular}{|m{0.04\textwidth}|m{0.14\textwidth}|m{0.12\textwidth}|>{\centering\arraybackslash}m{0.18\textwidth}|>{\centering\arraybackslash}m{0.18\textwidth}|>{\centering\arraybackslash}m{0.18\textwidth}|}
        \hline
        \textbf{No.} & \textbf{City} & \textbf{Routes} & \texttt{test-data-a.csv} & \texttt{test-data-b.csv} & \texttt{test-data-c.csv}\\
        \hline
        54 & Atlanta & GTFS \cite{marta-st} & 5.29 \% & 8.60 \% & 7.54 \% \\
        \hline
        55 & Miami & GTFS \cite{mdt-st} & 8.89 \% & 9.90 \% & 12.30 \% \\
        \hline
        56 & Tampa & GTFS \cite{hart-st} & 3.91 \% & 0.00 \% & 0.00 \% \\
        \hline
        57 & Louisville & GTFS \cite{tarc-st} & 9.83 \% & 18.42 \% & 15.96 \% \\
        \hline
        58 & Nashville & GTFS \cite{nas-mta-st} & 10.34 \% & 10.35 \% & 13.87 \% \\
        \hline
        59 & \multicolumn{2}{|l|}{\textbf{Average (Unweighted)}} &  \textbf{6.87 \%} & \textbf{9.45 \%} & \textbf{9.93 \%} \\
        \hline
        60 & \multicolumn{2}{|l|}{\textbf{Average (Weighted)}} & \textbf{7.53 \%} & \textbf{9.73 \%} & \textbf{11.08 \%} \\
        \hline
        \end{tabular}
    }
\caption{US South Percentages of \textbf{latitude -- longitude} 25m Outside GTFS Routes (WGS84)}
\label{tab:spl-pnt-us-suth}
\end{center}
\end{table}

\begin{table}[H]
\begin{center}
    \resizebox{\textwidth}{!}{
        \begin{tabular}{|m{0.04\textwidth}|m{0.14\textwidth}|m{0.12\textwidth}|>{\centering\arraybackslash}m{0.18\textwidth}|>{\centering\arraybackslash}m{0.18\textwidth}|>{\centering\arraybackslash}m{0.18\textwidth}|}
        \hline
        \textbf{No.} & \textbf{City} & \textbf{Routes} & \texttt{test-data-a.csv} & \texttt{test-data-b.csv} & \texttt{test-data-c.csv}\\
        \hline
        61 & \textit{Minneapolis} & GTFS \cite{mn-metro-st} & 9.08 \% & 12.35 \% & 10.91 \% \\
        \hline
        62 & St. Louis & GTFS \cite{stl-metro-st} & 0.60 \% & 4.22 \% & 1.42 \% \\
        \hline
        63 & Madison & GTFS \cite{msn-metro-st} & 13.81 \% & 14.59 \% & 16.95 \% \\
        \hline
        64 & Columbus & GTFS \cite{cota-st} & 11.35 \% & 6.09 \% & 6.24 \% \\
        \hline
        65 & Des Moines & GTFS \cite{dart-st} & 4.98 \% & 4.13 \% & 2.49 \% \\
        \hline
        66 & \multicolumn{2}{|l|}{\textbf{Average (Unweighted)}} &  \textbf{7.96 \%} & \textbf{8.28 \%} & \textbf{7.60 \%} \\
        \hline
        67 & \multicolumn{2}{|l|}{\textbf{Average (Weighted)}} & \textbf{9.47 \%} & \textbf{12.29 \%} & \textbf{10.63 \%} \\
        \hline
        \end{tabular}
    }
\caption{US Central Percentages of \textbf{latitude -- longitude} 25m Outside GTFS Routes (WGS84)}
\label{tab:spl-pnt-us-cent}
\end{center}
\end{table}

\begin{table}[H]
\begin{center}
    \resizebox{\textwidth}{!}{
        \begin{tabular}{|m{0.04\textwidth}|m{0.14\textwidth}|m{0.12\textwidth}|>{\centering\arraybackslash}m{0.18\textwidth}|>{\centering\arraybackslash}m{0.18\textwidth}|>{\centering\arraybackslash}m{0.18\textwidth}|}
        \hline
        \textbf{No.} & \textbf{City} & \textbf{Routes} & \texttt{test-data-a.csv} & \texttt{test-data-b.csv} & \texttt{test-data-c.csv}\\
        \hline
        68 & Denver & GTFS \cite{rtd-st} & 2.85 \% & 4.16 \% & 4.14 \% \\
        \hline
        69 & Phoenix & GTFS \cite{phx-metro-st} & 10.26 \% & 11.26 \% & 12.03 \% \\
        \hline
        70 & San Antonio & GTFS \cite{via-metro-st} & 1.32 \% & 3.13 \% & 2.25 \% \\
        \hline
        71 & Austin & GTFS \cite{cap-st} & 0.74 \% & 0.11 \% & 0.25 \% \\
        \hline
        72 & Billings & GTFS \cite{met-st} & ---\textemdash\space \% & ---\textemdash\space \% & ---\textemdash\space \% \\
        \hline
        73 & \multicolumn{2}{|l|}{\textbf{Average (Unweighted)}} &  \textbf{3.79 \%} & \textbf{4.66 \%} & \textbf{4.67 \%} \\
        \hline
        74 & \multicolumn{2}{|l|}{\textbf{Average (Weighted)}} & \textbf{8.34 \%} & \textbf{9.99 \%} & \textbf{10.85 \%} \\
        \hline
        \end{tabular}
    }
\caption{US Mountain Percentages of \textbf{latitude -- longitude} 25m Outside GTFS Routes (WGS84)}
\label{tab:spl-pnt-us-mntn}
\end{center}
\end{table}

\begin{table}[H]
\begin{center}
    \resizebox{\textwidth}{!}{
        \begin{tabular}{|m{0.04\textwidth}|m{0.14\textwidth}|m{0.12\textwidth}|>{\centering\arraybackslash}m{0.18\textwidth}|>{\centering\arraybackslash}m{0.18\textwidth}|>{\centering\arraybackslash}m{0.18\textwidth}|}
        \hline
        \textbf{No.} & \textbf{City} & \textbf{Routes} & \texttt{test-data-a.csv} & \texttt{test-data-b.csv} & \texttt{test-data-c.csv}\\
        \hline
        75 & Montréal & GTFS \cite{stm-st} & 1.71 \% & 1.93\% & 2.83\% \\
        \hline
        76 & York & GTFS \cite{yrt} & 3.13 \% & 3.04 \% & 3.85 \% \\
        \hline
        77 & Hamilton & GTFS \cite{hsr-st} & 17.54 \% & 14.95 \% & 16.41 \% \\
        \hline
        78 & Halifax & GTFS \cite{hat-st} & 1.09 \% & 1.08 \% & 0.89 \% \\
        \hline
        79 & Thunder Bay & GTFS \cite{tbt-st} & 4.20 \% & 1.72 \% & 2.26 \% \\
        \hline
        80 & \multicolumn{2}{|l|}{\textbf{Average (Unweighted)}} &  \textbf{5.53 \%} & \textbf{4.54 \%} & \textbf{5.25 \%} \\
        \hline
        81 & \multicolumn{2}{|l|}{\textbf{Average (Weighted)}} & \textbf{2.45 \%} & \textbf{2.40 \%} & \textbf{3.31 \%} \\
        \hline
        \end{tabular}
    }
\caption{Canada East Percentages of \textbf{latitude -- longitude} 25m Outside GTFS Routes (WGS84)}
\label{tab:spl-pnt-ca-east}
\end{center}
\end{table}

\begin{table}[H]
\begin{center}
    \resizebox{\textwidth}{!}{
        \begin{tabular}{|m{0.04\textwidth}|m{0.14\textwidth}|m{0.12\textwidth}|>{\centering\arraybackslash}m{0.18\textwidth}|>{\centering\arraybackslash}m{0.18\textwidth}|>{\centering\arraybackslash}m{0.18\textwidth}|}
        \hline
        \textbf{No.} & \textbf{City} & \textbf{Routes} & \texttt{test-data-a.csv} & \texttt{test-data-b.csv} & \texttt{test-data-c.csv}\\
        \hline
        82 & Vancouver & GTFS \cite{tl-bc-st} & 3.24 \% & 3.84 \% & 3.77 \% \\
        \hline
        83 & Calgary & GTFS \cite{cal-st} & 9.82 \% & 0.03 \% & 0.25 \% \\
        \hline
        84 & Edmonton & GTFS \cite{ets-st} & 0.98 \% & 1.34 \% & 4.41 \% \\
        \hline
        85 & Saskatoon & GTFS \cite{sas-st} & 4.38 \% & 5.58 \% & 9.47 \% \\
        \hline
        86 & \multicolumn{2}{|l|}{\textbf{Average (Unweighted)}} &  \textbf{4.60 \%} & \textbf{2.70 \%} & \textbf{4.47 \%} \\
        \hline
        87 & \multicolumn{2}{|l|}{\textbf{Average (Weighted)}} & \textbf{2.14 \%} & \textbf{2.40 \%} & \textbf{4.36 \%} \\
        \hline
        \end{tabular}
    }
\caption{Canada West Percentages of \textbf{latitude -- longitude} 25m Outside GTFS Routes (WGS84)}
\label{tab:spl-pnt-ca-west}
\end{center}
\end{table}

\begin{table}[H]
\begin{center}
    \resizebox{\textwidth}{!}{
        \begin{tabular}{|m{0.04\textwidth}|m{0.14\textwidth}|m{0.12\textwidth}|>{\centering\arraybackslash}m{0.18\textwidth}|>{\centering\arraybackslash}m{0.18\textwidth}|>{\centering\arraybackslash}m{0.18\textwidth}|}
        \hline
        \textbf{No.} & \textbf{City} & \textbf{Routes} & \texttt{test-data-a.csv} & \texttt{test-data-b.csv} & \texttt{test-data-c.csv}\\
        \hline
        88 & \textit{Amsterdam} & GTFS \cite{ov-st} & 3.68 \% & 3.7\% & 3.39 \% \\
        \hline
        89 & Stockholm & ---\textemdash\space & ---\textemdash\space \% & ---\textemdash\space \% & ---\textemdash\space \% \\
        \hline
        90 & Helsinki & GTFS \cite{hsl-st} & 2.46 \% & 2.95 \% & 1.79 \% \\
        \hline
        91 & \textit{Dublin} & GTFS \cite{nta-st} & 10.61 \% & 2.90 \% & 3.61 \% \\
        \hline
        92 & Rome & GTFS \cite{atac} & 7.28 \% & 6.36 \% & 1.45 \% \\
        \hline
        93 & \multicolumn{2}{|l|}{\textbf{Average (Unweighted)}} &  \textbf{6.01 \%} & \textbf{3.98 \%} & \textbf{2.56 \%} \\
        \hline
        94 & \multicolumn{2}{|l|}{\textbf{Average (Weighted)}} & \textbf{4.76 \%} & \textbf{3.44 \%} & \textbf{3.05 \%} \\
        \hline
        \end{tabular}
    }
\caption{EU West Percentages of \textbf{latitude -- longitude} 25m Outside GTFS Routes (WGS84)}
\label{tab:spl-pnt-eu-west}
\end{center}
\end{table}

\begin{table}[H]
\begin{center}
    \resizebox{\textwidth}{!}{
        \begin{tabular}{|m{0.04\textwidth}|m{0.14\textwidth}|m{0.12\textwidth}|>{\centering\arraybackslash}m{0.18\textwidth}|>{\centering\arraybackslash}m{0.18\textwidth}|>{\centering\arraybackslash}m{0.18\textwidth}|}
        \hline
        \textbf{No.} & \textbf{City} & \textbf{Routes} & \texttt{test-data-a.csv} & \texttt{test-data-b.csv} & \texttt{test-data-c.csv}\\
        \hline
        95 & Warsaw & ---\textemdash\space & ---\textemdash\space \% & ---\textemdash\space \% & ---\textemdash\space \% \\
        \hline
        96 & Kraków & ---\textemdash\space & ---\textemdash\space \% & ---\textemdash\space \% & ---\textemdash\space \% \\
        \hline
        97 & Gdańsk & ---\textemdash\space & ---\textemdash\space \% & ---\textemdash\space \% & ---\textemdash\space \% \\
        \hline
        98 & Prague & GTFS \cite{pid-st} & 2.72 \% & 6.00 \% & 6.40 \% \\
        \hline
        99 & \multicolumn{2}{|l|}{\textbf{Average (Unweighted)}} &  \textbf{2.72 \%} & \textbf{6.00 \%} & \textbf{6.40 \%} \\
        \hline
        100 & \multicolumn{2}{|l|}{\textbf{Average (Weighted)}} & \textbf{2.72 \%} & \textbf{6.00 \%} & \textbf{6.40 \%} \\
        \hline
        \end{tabular}
    }
\caption{EU Central Percentages of \textbf{latitude -- longitude} 25m Outside GTFS Routes (WGS84)}
\label{tab:spl-pnt-eu-cent}
\end{center}
\end{table}

\begin{table}[H]
\begin{center}
    \resizebox{\textwidth}{!}{
        \begin{tabular}{|m{0.04\textwidth}|m{0.14\textwidth}|m{0.12\textwidth}|>{\centering\arraybackslash}m{0.18\textwidth}|>{\centering\arraybackslash}m{0.18\textwidth}|>{\centering\arraybackslash}m{0.18\textwidth}|}
        \hline
        \textbf{No.} & \textbf{City} & \textbf{Routes} & \texttt{test-data-a.csv} & \texttt{test-data-b.csv} & \texttt{test-data-c.csv}\\
        \hline
        95 & Sydney & GTFS \cite{nsw-st} & 1.48 \% & 1.63 \% & 1.83 \% \\
        \hline
        96 & Brisbane & GTFS \cite{tl-ql-st} & 4.05 \% & 5.89 \% & 6.41 \% \\
        \hline
        97 & Adelaide & GTFS \cite{adl-metro-st} & 9.77 \% & 7.21 \% & 6.96 \% \\
        \hline
        98 & Auckland & GTFS \cite{at-st} & 7.67 \% & 7.70 \% & 7.67 \% \\
        \hline
        99 & Christchurch & GTFS \cite{ecan-st} & 1.83 \% & 2.10 \% & 2.03 \% \\
        \hline
        100 & \multicolumn{2}{|l|}{\textbf{Average (Unweighted)}} &  \textbf{4.96 \%} & \textbf{4.91 \%} & \textbf{4.98 \%} \\
        \hline
        101 & \multicolumn{2}{|l|}{\textbf{Average (Weighted)}} & \textbf{5.79 \%} & \textbf{6.01 \%} & \textbf{6.45 \%} \\
        \hline
        \end{tabular}
    }
\caption{Oceania Percentages of \textbf{latitude -- longitude} 25m Outside GTFS Routes (WGS84)}
\label{tab:spl-pnt-oceania}
\end{center}
\end{table}

\begin{table}[H]
\begin{center}
    \resizebox{\textwidth}{!}{
        \begin{tabular}{|m{0.04\textwidth}|m{0.14\textwidth}|m{0.12\textwidth}|>{\centering\arraybackslash}m{0.18\textwidth}|>{\centering\arraybackslash}m{0.18\textwidth}|>{\centering\arraybackslash}m{0.18\textwidth}|}
        \hline
        \textbf{No.} & \textbf{City} & \textbf{Routes} & \texttt{test-data-a.csv} & \texttt{test-data-b.csv} & \texttt{test-data-c.csv}\\
        \hline
        102 & Delhi & ---\textemdash\space & ---\textemdash\space \% & ---\textemdash\space \% & ---\textemdash\space \% \\
        \hline
        103 & \multicolumn{2}{|l|}{\textbf{Average (Unweighted)}} &  \textbf{---\textemdash\space \%} & \textbf{---\textemdash\space \%} & \textbf{---\textemdash\space \%} \\
        \hline
        104 & \multicolumn{2}{|l|}{\textbf{Average (Weighted)}} &  \textbf{---\textemdash\space \%} & \textbf{---\textemdash\space \%} & \textbf{---\textemdash\space \%} \\
        \hline
        \end{tabular}
    }
\caption{Asia Percentages of \textbf{latitude -- longitude} 25m Outside GTFS Routes (WGS84)}
\label{tab:spl-pnt-asia}
\end{center}
\end{table}

\subsubsection{Temporal Contiguity}
\label{sub-sec:tmp-cnt}

Continuous validity and consistency was globally tested throughout all validation periods. Exhibited in Table \ref{tab:tmp-cnt}, test numbers 105 and 106 did not find any invalid values in the \textbf{datetime} field. Test number 107 found the average (unweighted) proportion of time that no observations were measured for longer than 1 minute was 2.75\%. Test number 108 found the average (unweighted) longest single interval that no observations were measured was 4 minutes, 24 seconds. Also, in test 109, the average (unweighted) elapsed time from the first observation to the last was 23 hours and 58 minutes within the context of a 24-hour maximum time horizon, exhibited in Figure \ref{fig:tmp-cnt}.

\begin{table}[H]
\begin{center}
    \resizebox{\textwidth}{!}{
        \begin{tabular}{|m{0.04\textwidth}|m{0.28\textwidth}|>{\centering\arraybackslash}m{0.18\textwidth}|>{\centering\arraybackslash}m{0.18\textwidth}|>{\centering\arraybackslash}m{0.18\textwidth}|}
        \hline
        \textbf{No.} & \textbf{datetime} & \texttt{test-data-a.csv} & \texttt{test-data-b.csv} & \texttt{test-data-c.csv}\\
        \hline
        105 & Zero Values & 0.00 \% & 0.00 \% & 0.00 \% \\
        \hline
        106 & Negative Values & 0.00 \% & 0.00 \% & 0.00 \% \\
        \hline
        107 & Downtime ($> 1$ min.) & 2.16 \% & 2.56 \% & 3.53 \% \\
        \hline
        108 & Downtime Max. Interval & 3 min. 57 sec. & 5 min. 21 sec. & 3 min. 56 sec. \\
        \hline
        109 & Elapsed Time & 24 hrs. 0 min. & 23 hrs. 57 min. & 23 hrs. 59 min. \\
        \hline
        \end{tabular}
    }
\caption{Global Temporal Contiguity in \textbf{datetime} Field}
\label{tab:tmp-cnt}
\end{center}
\end{table}

\begin{figure}[H]
\centering
    \begin{tikzpicture}
        \begin{axis}[
            ybar,
            bar width=3pt,
            width=\textwidth,
            height=5cm,
            legend style={
                at={(0.98, 0.05)},
                anchor=south east,
                font=\small,
                fill opacity=0.90,
                text opacity=1,
                legend image post style={yshift=-2pt}
            },  
            ylabel={Number of Observations},
            ylabel style={font=\footnotesize},
            xlabel={Hour},
            xlabel style={font=\footnotesize},
            symbolic x coords=  {1,2,3,4,5,6,7,8,9,10,11,12,13,14,15,16,17,18,19,20,21,22,23,24},
            xtick=data,
            enlarge x limits={0.03},
            nodes near coords align={vertical},
            ymin=0,ymax=10500, 
            ylabel near ticks,
            xlabel near ticks,
            xtick style={draw=none},
            ytick style={draw=none},
            scaled y ticks=false,
            yticklabel style={/pgf/number format/fixed},
            ytick={0,2000,...,10000},
            extra y ticks={1000,3000,5000,7000,9000},
            extra y tick style={grid=major, tick style={draw=none}, grid style={dashed}},
            legend image code/.code={ 
                \draw plot coordinates {(0, 4pt)};
            }
        ]
        
        \addplot[fill=yellow!60!black, color=yellow!60!black] coordinates {
            (1,6954) (2,8322) (3,8696) (4,8538) (5,9358) (6,8459) (7,7382) (8,7892) (9,7433) (10,7529) (11,7683) (12,7623) (13,6817) (14,8383) (15,8682) (16,9237) (17,8812) (18,8640) (19,8355) (20,7028) (21,6296) (22,7017) (23,8609) (24,8492)
        };
        
        \addplot[fill=orange!80!black, color=orange!80!black] coordinates {
            (1,8080) (2,7167) (3,8965) (4,8778) (5,8615) (6,8917) (7,9323) (8,7226) (9,9535) (10,9203) (11,5940) (12,9353) (13,9411) (14,7317) (15,6700) (16,7349) (17,7747) (18,6241) (19,8239) (20,8521) (21,9094) (22,8789) (23,9471) (24,7430)
        };
        
        \addplot[fill=blue!80!black, color=blue!80!black] coordinates {
            (1,8413) (2,9752) (3,9225) (4,9709) (5,9291) (6,9744) (7,8125) (8,7653) (9,9686) (10,9087) (11,8749) (12,8759) (13,9831) (14,8852) (15,6565) (16,7413) (17,9593) (18,6938) (19,9085) (20,9841) (21,10002) (22,9804) (23,7172) (24,6040)
        };
        
        \legend{
            \texttt{test-data-a.csv}, \texttt{test-data-b.csv}, \texttt{test-data-c.csv}
        }
        \end{axis}
    \end{tikzpicture}
\caption{Global Temporal Contiguity in \textbf{datetime} Field}
\label{fig:tmp-cnt}
\end{figure}

\subsubsection{Expected Duration}
\label{sub-sec:dur-exp}

The global average idling duration was as expected throughout all validation periods. Exhibited in Table \ref{tab:dur-exp}, test numbers 110 and 111 did not find any invalid values in the \textbf{duration} field. Test number 112 found when the minimum idling duration was adjusted to a commonly used policy definition longer than 5 minutes \cite{epa2006}, the proportion of idling time denominated by total operational time was 38.74\%. This is broadly consistent and within the 30\% to 44\% range from existing studies, previously mentioned in \nameref{sec:back}. Figure \ref{fig:dur-exp} exhibits this comparison, as well as when the minimum idling tolerance was unadjusted according to the parameterized definition $h = 1$ minute. \\ 

When the minimum idling duration in test number 113 was unadjusted according to $h = 1$ minute, the average (unweighted) proportion of idling increased to 56.96\%. Although approximately 18\% of the average proportion of idling increased as the minimum idling tolerance decreased from 5 minutes to 1 minute, it is concluded that the \textbf{duration} field is still reliable and valid given its combined consistency with existing studies when adjusted to common policy definitions of idling. The material difference between tests 112 and 113 is likely not a measurement error, but rather a unique feature of this data that warrants further investigation in downstream analyses.

\begin{table}[H]
\begin{center}
    \resizebox{\textwidth}{!}{
        \begin{tabular}{|m{0.04\textwidth}|m{0.28\textwidth}|>{\centering\arraybackslash}m{0.18\textwidth}|>{\centering\arraybackslash}m{0.18\textwidth}|>{\centering\arraybackslash}m{0.18\textwidth}|}
        \hline
        \textbf{No.} & \textbf{duration} & \texttt{test-data-a.csv} & \texttt{test-data-b.csv} & \texttt{test-data-c.csv}\\
        \hline
        110 & Zero Values & 0.00 \% & 0.00 \% & 0.00 \% \\
        \hline
        111 & Negative Values & 0.00 \% & 0.00 \% & 0.00 \% \\
        \hline
        112 & Avg. Idle Adj. ($>5$ min.) & 34.32 \% & 40.32 \% & 41.60 \% \\
        \hline
        113 & Avg. Idle ($>h=1$ min.) & 52.48 \% & 58.62 \% & 59.78 \% \\
        \hline
        \end{tabular}
    }
\caption{Global Expected Idling Duration Percentages in \textbf{duration} Field}
\label{tab:dur-exp}
\end{center}
\end{table}

\vspace{-12pt}
\begin{figure}[H]
\centering    
    \begin{tikzpicture}
    \begin{axis}[
        xlabel={Proportion of Idling},
        symbolic y coords={Avg. Idle Adj.,Avg. Idle}, 
        ytick={Avg. Idle Adj.,Avg. Idle}, xtick={0, 30, 44, 100},
        xticklabels={0\%,30\%,44\%,100\%},
        xmin=0, xmax=100,
        height=6cm, 
        width=\textwidth-1.5cm,
        y=1.5cm,
        xmajorgrids=true, ymajorgrids=true, grid style=dashed,
        extra x ticks={48,62},
        extra x tick labels={48\%,62\%},
        extra x tick style={
            ticklabel pos=right,
            major tick length=0pt,
            xlabel style={anchor=south, yshift=5pt}
        }
    ]
    
    \addplot[
        color=red!60!black,
        mark=none,
        line width=1cm,
    ] coordinates {
        (53,Avg. Idle) 
        (60,Avg. Idle)
    };
    
    \addplot[
        color=green!60!black,
        mark=none,
        line width=1cm,
    ] coordinates {
        (34,Avg. Idle Adj.) 
        (42,Avg. Idle Adj.)
    };
    
    \node at (axis cs:57,Avg. Idle) [anchor=south, yshift=-32pt] {53\% \space 
     60\%};
    \node at (axis cs:38,Avg. Idle Adj.) [anchor=north, yshift=32pt] {34\% \space 42\%};
    
    \end{axis}
    \end{tikzpicture}
\caption{Global Expected Idling Duration Definitional Difference in \textbf{duration} Field}
\label{fig:dur-exp}
\end{figure}

\section{Limitations}
\label{sec:lim}

The geographic coverage of this data is constrained to regions with publicly available GTFS Realtime feeds, which introduces systemic bias toward regions with developed infrastructure, predominantly in North America, Europe, Oceania, and portions of Asia. Developing regions with substantial transit bus operations in Africa and South America remain underrepresented or absent.\\

Also, the data does not capture private operators, informal transit services, and agencies without digital tracking infrastructure. Vehicle-specific characteristics such as engine type, age, maintenance status affecting actual emissions are not captured and the system cannot distinguish between operational idling types such as traffic congestion, passenger boarding, mechanical failures.\\

In addition, the system's accuracy depends entirely on the frequency and positional precision of GTFS Realtime feeds that can vary across transit agencies. GPS accuracy limitations may result in false positives when buses are moving slowly or false negatives when buses idle during brief stops. This may fail to capture very brief idling events or misclassify very slow traffic movement.

\section{Ethics Statement}
\label{sec:ethics}
The authors confirm that this work does not involve human subjects, animal experiments, or data collected from social media platforms.

\section{CRediT Author Statement}
\textbf{Nicholas Kunz:} Conceptualization; Methodology; Software; Validation; Formal Analysis; Investigation; Data Curation; Visualization; Writing – Original Draft; Writing – Review \& Editing; Project Administration. \textbf{H. Oliver Gao:} Conceptualization; Supervision; Resources; Funding Acquisition; Writing – Review \& Editing: Project Administration.

\section{Acknowledgements}
We would like to thank the transit agencies for providing public developer support for GTFS and GTFS Realtime and those that accommodated special requests for enhanced permissions and support during testing and development.\\

Thank you Dr. Mengjie Han at MTC (San Francisco), Andrew Lowe and Dante Avery at Swiftly (L.A., Miami, Tampa), NTA (Dublin), and Trafiklab (Stockholm). Also, thank you to Dr. Graeme Troxell and Siva Visvesvaran for your encouragement and preliminary feedback.\\

This research did not receive any specific grant from funding agencies in the public, commercial, or not-for-profit sectors. This research was supported by Gao Labs and Cornell Systems Engineering.

\section{Declaration of Competing Interests}
The authors declare that they have no known competing financial interests or personal relationships that could have appeared to influence the work reported in this paper.

\section{Data Availability}
\label{sec:notes}

It is highly encouraged that this realtime data is indeed used in realtime applications and analyses, as well as with historical methods. In doing so, requires a functioning version of the software to be deployed. More information can be found at: \href{https://www.github.com/nickkunz/idling}{https://www.github.com/nickkunz/idling} \cite{kunz2024}. Updated versions of the \nameref{sub-sec:stc-fil} are available at: \href{https://doi.org/10.6084/m9.figshare.25224224}{https://doi.org/10.6084/m9.figshare.25224224} \cite{kunzgao2024}.

\section{Appendix}
\label{sub-sec:algo-info}
\begin{algorithm}[H]
\caption{GRD-TRT-BUF-4I}
\textbf{Input}: $r > 0, h > 0, m > 0$\\
\textbf{Output}: $\mathcal{Y} = \bigl\{[v_{(i,j)\,T}]: v \in \mathcal{Y} \bigl\}$
\begin{algorithmic}[1]
\STATE $i = 1$
\WHILE{$i$}
    \STATE $d_i \gets$ $\text{Feed}_i \gets$ Server 
    \IF{$i \geq T = t+h+2$}
        \STATE \textbf{delete} $d_{i+h+2}$ \hfill\COMMENT{\,5:\,limit buffer length}
        \STATE $A \gets d_0$, $B \gets d_h$, $C \gets d_{h+1}$
        \STATE $n(A^a \cap B^b) = A^a \cap B^b$
        \STATE $\mathcal{A} = \bigl\{[x_{(i,\,j=1,\ldots,5)\,t}]: x \in A^n \bigl\}$, 
            $\mathcal{B} = \bigl\{[y_{(i,\,j=1,\ldots,5)\,t+h}]: y \in B^n \bigl\}$
        \STATE $\mathcal{H} \gets \mathcal{H}_i = \mathcal{A} \cap \mathcal{B}$, $\mathcal{C} = \bigl\{[z_{(i,\,j=1,\ldots,5)\,t+h+1}]: z \in C \bigl\}$
        \FOR {$i \in \mathcal{H}$}
            \IF{$i \notin k_i$}
                \STATE $k_i \gets 0$
            \ENDIF
            \IF{$i \notin \mathcal{C}$}
                \STATE $k_i = k_i + 1$
            \ELSE
                \STATE $k_i = 0$
            \ENDIF
            \IF{$k_i \geq m$}
                \STATE \textbf{delete} $k_i$, $H_i$ \hfill\COMMENT{\,20:\,bound append length}
            \ENDIF
        \ENDFOR
        \STATE $\mathcal{Y} = \mathcal{H} \cap \mathcal{C}$
        \STATE \textbf{return} $\mathcal{Y}$
        \STATE $\text{Store} \gets \mathcal{Y}$
    \ELSE
        \STATE \textbf{continue}
    \ENDIF
    \STATE $i=i+1$
    \STATE \textbf{pause} $r$ \text{sec} \hfill\COMMENT{\,30:\,limit request rate}
\ENDWHILE
\end{algorithmic}
\end{algorithm}
\label{alg:grd-trt-buf-4i}

\newpage
\bibliographystyle{unsrt}

\end{document}